\documentclass[useAMS,usenatbib,usegraphicx]{mn2e}
\usepackage{amsmath,mathtools,amsfonts,amssymb,float}
\usepackage[usenames,dvipsnames]{xcolor}
\usepackage{pdfpages}
\usepackage{soul}

\def\Msun{\hbox{$\rm\, M_{\odot}$}}

\title [Large-scale galactic conformity] {On the evidence for large-scale galactic conformity in the local Universe}

\author[L. Sin, S. Lilly, and B. Henriques]  
{Larry P. T. Sin$^{1}$\thanks{E-mail: sinp@phys.ethz.ch},
Simon J. Lilly$^{1}$, Bruno M. B. Henriques$^{1}$\vspace{0.4cm}\\
  {}$^{1}$Institute for Astronomy, Department of Physics, ETH Z\"urich, CH-8093 Z\"urich, Switzerland\\}
 
\begin{document}

\date{Accepted 2017 June 30. Received 2017 June 26; in original form 2017 February 27}

\volume{471}
\pagerange{1192--1207} \pubyear{2017}

\maketitle

\label{firstpage}

\begin{abstract}
We re-examine the observational evidence for large-scale (4 Mpc) galactic conformity in the local Universe, as presented in \citet{Kauffmann2013}. We show that a number of methodological features of their analysis act to produce a misleadingly high amplitude of the conformity signal. These include a weighting in favour of central galaxies in very high-density regions, the likely misclassification of satellite galaxies as centrals in the same high-density regions, and the use of medians to characterize bimodal distributions. We show that the large-scale conformity signal in Kauffmann et al. clearly originates from a very small number of central galaxies in the vicinity of just a few very massive clusters, whose effect is strongly amplified by the methodological issues that we have identified. Some of these \lq centrals\rq\ are likely misclassified satellites, but some may be genuine centrals showing a real conformity effect. Regardless, this analysis suggests that conformity on 4 Mpc scales is best viewed as a relatively short-range effect (at the virial radius) associated with these very large neighbouring haloes, rather than a very long-range effect (at tens of virial radii) associated with the relatively low-mass haloes that host the nominal central galaxies in the analysis. A mock catalogue constructed from a recent semi-analytic model shows very similar conformity effects to the data when analysed in the same way, suggesting that there is no need to introduce new physical processes to explain galactic conformity on 4 Mpc scales.
\end{abstract}


\section{Introduction}
Galaxies are known to populate a broadly bimodal distribution in their star-formation rates \citep{Kauffmann2003, Baldry2004, Brinchmann2004}. On the one hand, there is a population of star-forming galaxies in which the star-formation rate closely follows the stellar mass, producing a so-called \lq Main Sequence\rq\ in which the specific star-formation rate (sSFR) has only a weak variation with stellar mass, and a dispersion of only about a factor of 2 \citep{Brinchmann2004, Daddi2007, Elbaz2007, Noeske2007, Salim2007}. On the other hand, there is also a population of galaxies in which the rate of star-formation is suppressed by one or two orders of magnitude relative to the Main Sequence. Such galaxies are evolving passively. We will refer to these two populations, split by sSFR, as \lq star-forming\rq\ and \lq passive\rq\ respectively.

Understanding the process(es) by which galaxies transition from the star-forming to the passive population, a transition that is often called \lq quenching\rq, is a major goal for the study of galaxy evolution. It is clear that both the mass and the environment of a galaxy play a role. For instance, the fraction $f_{\mathrm{Q}}$ of galaxies that are quenched in the local SDSS sample is a separable function in terms of the stellar mass of the galaxies and of the local density of galaxies around them \citep{Peng2010}. Peng et al. coined the phrases \lq mass-quenching\rq\ and \lq environment-quenching\rq\ to describe these two drivers of quenching. Most galaxies in high-density environments are satellite galaxies, i.e. galaxies orbiting within the dark matter halo of another, more massive galaxy, called the central galaxy. Environment-quenching is dominated by the quenching of satellite galaxies \citep{Peng2012}. A satellite-quenching efficiency $\epsilon_{\mathrm{sat}}$, defined as the excess probability that a satellite is quenched, relative to if it were a central of the same stellar mass, is strikingly independent of its stellar mass \citep{vdBosch2008, Peng2012}, which is why $f_{\mathrm{Q}}$ appears separable in stellar mass and density.

One difficulty in moving towards a physical understanding of quenching is in identifying which \lq mass\rq\ and which \lq environment\rq\ are really driving it. In the $\Lambda$CDM paradigm, galaxies form and evolve at the bottom of the potential wells of collapsed dark matter haloes. For central galaxies, there is a tight correlation between the stellar mass, the dark matter halo mass, and even the mass of the central supermassive black hole, and all three of these have been claimed as the driver of the highly mass-dependent mass-quenching process (e.g. \citealt{Baldry2006, Bower2006, Croton2006, Bluck2014, Woo2015, Henriques2016}). Similarly with environment, many mechanisms have been proposed for the quenching of star-formation in satellites, including ram-pressure stripping, tidal stripping of gas, the disruption of the fuel supply (\lq strangulation\rq), and the effect of close encounters between galaxies (\lq harassment\rq)(e.g. \citealt{Gunn1976, Larson1980, Moore1996, Abadi1999}).

However, for about a decade there have been indications that the processes that quench centrals and satellites may be closely linked. In particular, \citet{Weinmann2006} found that, at a fixed halo mass, passive centrals tend to have passive satellites, and star-forming centrals tend to have star-forming satellites. They named this correlation \lq galactic conformity\rq. 

It might be thought that conformity would arise if both the quenching of centrals and satellites were independently affected by the halo mass, since clearly higher mass haloes would be more likely to contain both quenched centrals and quenched satellites. However, if the samples of satellites and centrals are studied at a fixed halo mass, or with samples that are carefully matched in halo mass, as in \citet{Weinmann2006}, then the conformity signal from such independent effects should disappear. The persistence of conformity in halo-mass-matched samples is a clear indication that the evolution of star formation within galaxies is influenced by properties beyond halo mass. This has been discussed in detail by \citet{Knobel2015}, who analysed the conformity signal in SDSS groups and matched no less than five parameters, namely the halo mass, the normalized group-centric distance, the local density, the stellar mass of the central, and the stellar mass of the satellite. They showed that, even after matching these five parameters, there was still a strong conformity signal in the sense that the $\epsilon_{\mathrm{sat}}$ for satellites around quenched centrals was 2.5 times higher than for satellites around star-forming centrals.

The existence of conformity between centrals and their satellites requires either that the quenching of satellites is to some degree consequent on the quenching of the central (or vice versa), or that the quenching of both is being driven, across the halo, by another parameter which was not \lq matched\rq\ in the analysis (see \citealt{Knobel2015} for discussion). One possibility is effects that are linked to the assembly history of the halo. \citet{Yang2006} found in the low-redshift Universe that at fixed halo mass, the bias of galaxy groups decreases as the SFR of the central galaxy increases, while \citet{Gao2005} had earlier found within the Millennium simulation that at fixed halo mass, haloes that formed at earlier times also tend to be more biased (i.e. strongly clustered) than haloes that formed later, i.e. haloes that formed earlier might be expected to have older stellar populations.

A surprising development was the work of \citet[][hereafter K13]{Kauffmann2013}, who presented observational evidence for a strong conformity signal extending out to very large distances. In particular, they showed evidence (see their fig. 2) that, around centrals with stellar masses $10^{10}\Msun < M_*< 10^{10.5}\Msun$, a strong conformity signal extends out to 4 Mpc, i.e. of order ten times beyond the virial radii of the haloes that host these relatively low-mass central galaxies. Indeed, to first order, there is little variation in the strength of conformity with distance for these centrals. As well as the scale, the amplitude of the effect was also surprising: at distances of 3 Mpc from low-sSFR centrals, a suppression by a factor of 2 was seen in the sSFR distribution of neighbouring galaxies.

Taken at face value, this suggests that distinct haloes with no direct physical relation somehow share a common evolutionary path. This could arise from large-scale causal effects operating on super-halo scales (e.g. from AGN feedback, see \citealt{Kauffmann2015}), going against a commonly held assumption that the properties of the halo completely govern the properties of the galaxies therein, and indicating that a major effect is missing from our current understanding of galaxy formation and evolution.

Alternatively, it could arise from the fact that parameters which could be producing conformity within a single halo, such as the assembly history of haloes, or halo concentration, will be correlated on scales of 10 Mpc \citep[see][]{Hearin2015, Paranjape2015, Hearin2016}. However, studies arguing that large-scale conformity arises, via biasing, i.e. from the spatial correlation of one-halo effects, were not able to account for the strength of the effect presented in K13, although the last cited claimed that they were qualitatively similar. In semi-analytic models, which should in principle include the relevant baryonic processes within haloes, the predicted strength of this signal is an order of magnitude weaker than observed (e.g. fig. 9 of K13).

Given the important implications of their results, we have examined the methodology and observational evidence that was presented in K13. Our goal is to assess the extent to which the K13 result can be considered as evidence for the existence of strong large-scale conformity on scales of 4 Mpc, and to try to identify in more detail the precise origin(s) of this strong signal. While the primary focus is on conformity, some of the methodological points will have wider interest.

This paper is organized as follows: In Section~\ref{sec:input_data}, we describe the observational and simulated data used in this work. In Section~\ref{sec:met_k13}, we present a detailed examination of the K13 methodology and results, and highlight some features which are cause for concern. In Section~\ref{sec:re_analysis}, we illustrate the effects that the highlighted features had on the final conformity result. In Section~\ref{sec:mock_comparison}, we compare the observational results to those obtained from semi-analytic models. In Section~\ref{sec:discussion}, we discuss the more general implications that our findings have on the existence of large-scale conformity. Finally in Section~\ref{sec:summary} we summarize our conclusions.

We use a $\Lambda\mathrm{CDM}$ cosmology with $\Omega_{\Lambda}= 0.7$, $\Omega_\mathrm{M} = 0.3$, and $H_0 = 70\;\mathrm{km\,s}^{-1}\,\mathrm{Mpc}^{-1}$. We use the dimensionless unit \lq dex\rq\ to denote the anti-logarithm in base 10. That is to say, a multiplicative difference by a factor of $10^n$ in linear space is equal to an additive difference of $n$ dex in logarithmic space. Throughout this work, 1-$\sigma$ statistical uncertainties are estimated via 100 iterations of bootstrap resampling.
\label{sec:intro}

\section{Input data and reproduction of the K13 result}
\label{sec:input_data}

\subsection{Observational data}
\label{subsec:obs_data}
In order to replicate the results from K13, we follow as closely as possible their sample selection. We use the galaxy sample presented in the New York University Value-Added Galaxy Catalogue\footnote{http://cosmo.nyu.edu/blanton/vagc/} \citep{Blanton2005}, which was constructed with data from Data Release 7 of the Sloan Digital Sky Survey \citep[SDSS DR7;][]{Abazajian2009}. Estimates of stellar masses and star-formation rates are an updated version of those derived in \citet{Brinchmann2004}\footnote{http://wwwmpa.mpa-garching.mpg.de/SDSS/DR7/}. From this catalogue, we select galaxies which were primary spectroscopic targets, which have redshifts within $0.017 < z < 0.03$, and which have stellar masses above $10^{9.25} \Msun$. These cuts result in a mass complete sample of 13,928 galaxies.

\subsection{Reproduction of K13 fig. 2}
\label{subsec:rep_k13}
Using the selected SDSS sample, we first try to reproduce the key result presented in K13, i.e. their fig. 2, following as closely as possible their own analysis. 

We first identify a set of \lq centrals\rq, defined as any galaxy with stellar mass $M_{*,i} > 5\times10^9 \Msun$ which has no other galaxy that is more massive than $M_{*,i}/2$ within a projected distance of $R_{\mathrm{proj}} = 500 \, \mathrm{kpc}$ and within a velocity difference of 
$c\Delta z = \,500 \, \mathrm{km\,s^{-1}}$. This selection was referred to as the \lq isolation criterion\rq\ by K13.

Then, for each central, the projected distances are calculated to all other nearby galaxies (henceforth referred to as \lq neighbours\rq) that lie within $R_{\mathrm{proj}}=4\,\mathrm{Mpc}$ and $c\Delta z=\pm\,500\,\mathrm{km\,s^{-1}}$ of the central. We note that the velocity criterion for neighbours was not stated in K13, but we adopt $\pm\,500\,\mathrm{km\,s^{-1}}$ for consistency with the isolation criterion.

\begin{figure}
\centering
\includegraphics[width=8.6cm]{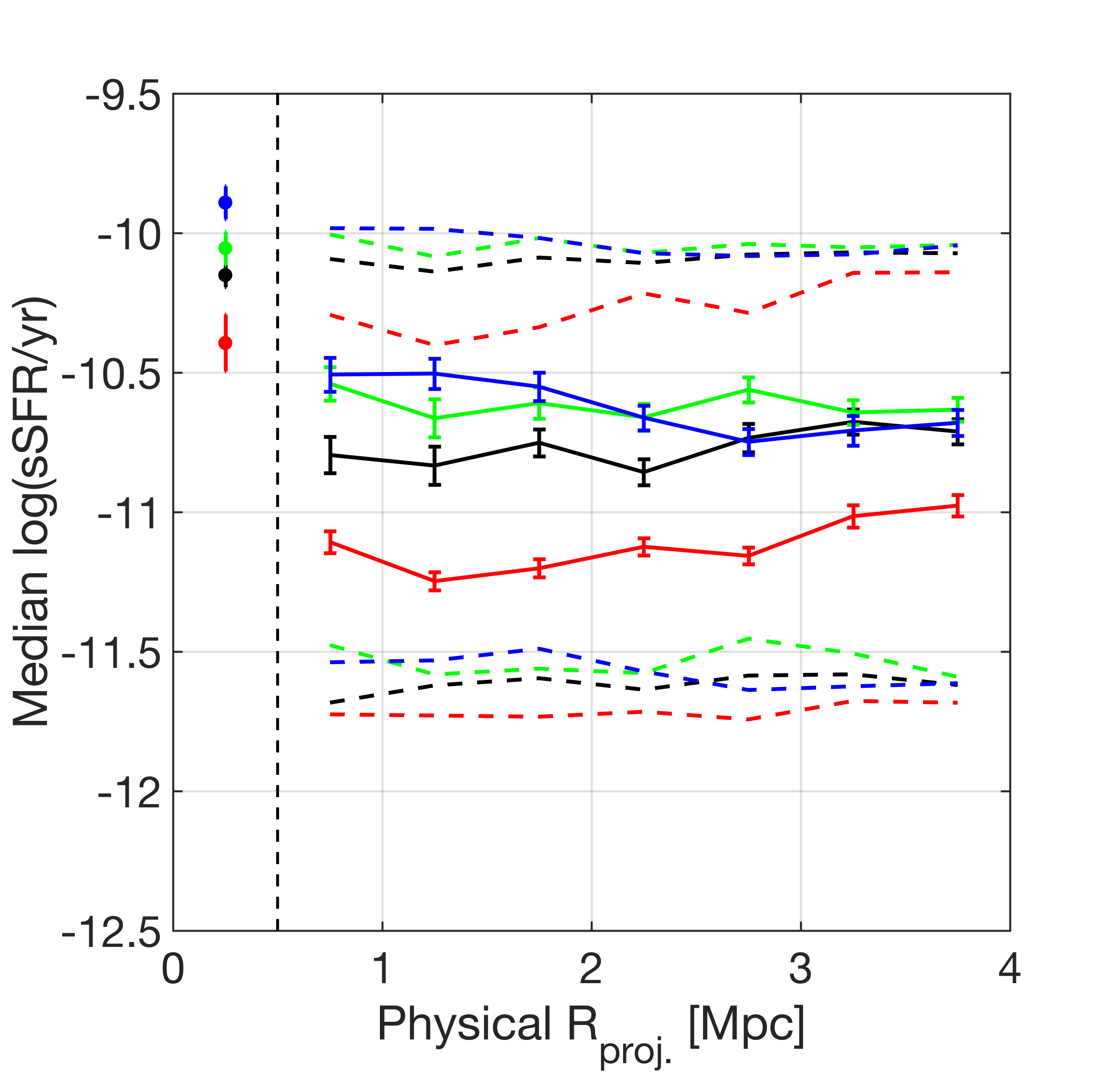}
\caption{Replication of fig. 2 in K13. The sSFR distribution of neighbours as a function of the projected distance-to-central, for centrals with masses between $10^{10}-10^{10.5} \Msun$. The red, black, green, and blue points and solid lines indicate the median sSFR for the neighbours of the centrals in the $0-25^{\mathrm{th}}$, $25-50^{\mathrm{th}}$, $50-75^{\mathrm{th}}$, and $75-100^{\mathrm{th}}$ percentile of sSFR, respectively. The dashed lines indicate the 25$^{\mathrm{th}}$ and 75$^{\mathrm{th}}$ percentiles of the neighbour sSFR distribution. The vertical dashed line at 500 kpc indicates the size of the aperture used for central selection. Within this radius, the masses of neighbour galaxies are biased low by the isolation criterion used to select the centrals. The sSFR of these inner neighbours are therefore biased high relative to the neighbours beyond 500 kpc.}
\label{fig:confor_kauff}
\end{figure}
While K13 presented results for three different mass ranges of centrals, we will focus only on centrals in the middle range $10^{10}-10^{10.5} \Msun$, for which the claimed conformity effect in K13 is strongest. Furthermore, in K13, the star-formation activity in the centrals was characterized not only by their sSFR (for both the fibre-aperture and a total estimate), but also by estimates of their HI gas fractions and HI-deficiencies that were derived from combinations of various observational parameters. Again, for the sake of brevity, we will focus only on the (total) sSFR. We are confident that the general conclusions presented in the current paper do not depend on these choices.

The set of centrals in our chosen mass range is then divided into quartiles by their sSFRs. For the centrals in each sSFR quartile, the distribution of sSFRs in the neighbour galaxies is then calculated as a function of the projected distance from the central, and represented using the median sSFR (as in K13). Neighbours are selected without regard to their stellar mass beyond the initial $M_{*,i} > 10^{9.25} \Msun$  selection.

Fig.~\ref{fig:confor_kauff} shows the result of this replication of the analysis in K13. It agrees very well with their analysis, and in particular with the bottom-right panel of fig. 2 in their work, which is the most directly comparable of their plots. A pronounced conformity-like correlation is seen, in that the neighbours of the centrals in the lowest sSFR-quartile (red line) have suppressed sSFRs relative to the neighbours of centrals in the other quartiles, extending all of the way out to 4 Mpc. Indeed, there is a general correlation between the average sSFR of the neighbours and the centrals over all four of the quartiles of central sSFR. We have checked that this signal is also present for centrals with masses down to $3\times10^9 \Msun$, and is also weakly present for those with masses above $10^{11} \Msun$, but that it is strongest for centrals between $10^{10} - 10^{10.5} \Msun$, as in K13.

While our results closely resemble those of K13, the depression of the red line relative to the others in our plot is somewhat weaker (overall by about 0.2 dex) than in K13. This difference can be due to a number of details. For instance, the neighbour star-formation activity in K13 is characterized by the sSFR evaluated within the SDSS fibre. On the other hand, we consistently use only the total sSFR for both centrals and neighbours in our work. 

These small differences aside, we do find that, by following the K13 methodology, the 4 Mpc neighbours of low-sSFR centrals do indeed have significantly lower sSFRs than average. Throughout this work, we will refer to this strong, long-range correlation between low-sSFR centrals and low-sSFR neighbours as \lq the signal\rq. The authors of K13 interpreted this signal as evidence in favour of the existence of conformity extending far beyond the haloes of the centrals in question, which would have virial radii $R_{\mathrm{vir}}\sim 250\,\mathrm{kpc}$ (as estimated in K13).

\subsection{Group catalogue}
\label{subsec:group_cat}

Although we will basically follow K13's identification of centrals using their isolation criterion, we will also make reference to the group membership of SDSS galaxies at some points in the discussion. We do so by making use of the Yang et al. SDSS DR7 group catalogue\footnote{http://gax.shao.ac.cn/data/Group.html}, the construction of which is described in \citet{Yang2007}.
	
The primary use of the group catalogue will be to identify which, if any, of the galaxies that were identified by K13 as centrals could in fact be satellites. Following the central selection criteria in \citet{Knobel2015}, we rank the members of a given group in mass, and also in their angular position relative to the median position of the group. We then define a central as being a galaxy within the top 10 percentile of its group in both mass and centrality. In the case where more than one galaxy satisfies these criteria, the most massive of these is assigned as the central if it is at least twice as massive as the second most massive; otherwise, the group is not assigned a central.

\subsection{Mock data}
\label{subsec:mock_cat}

In order to compare the results from observations against those predicted by galaxy formation models, we also use the semi-analytic model (SAM) of \citet[][hereafter H15]{Henriques2015}\footnote{http://galformod.mpa-garching.mpg.de/public/LGalaxies/}. This is the most recent major release of the so-called Munich models and was implemented on the Millennium dark matter simulations scaled to a Planck-year1 cosmology \citep{Planck2014}. Specifically, the cosmological parameters adopted are: $\sigma_8 = 0.829$, $H_0 = 67.3 \,\mathrm{km\,s^{-1}Mpc^{-1}}$, $\Omega_{\Lambda} = 0.685$, $\Omega_{\mathrm{M}} = 0.315$, $\Omega_{\mathrm{b}} = 0.0487$ $(f_{\mathrm{b}} = 0.155)$ and 
$n = 0.96$. We use the galaxy catalogue based on the Millennium simulation since it has a larger volume (meaning better statistics for satellite galaxies), and H15 showed that its properties converge with those in the catalogue based on the higher-resolution Millennium-II down to our low-mass limit ($M_*=10^{9.25} \Msun$). Both the Millennium and Millennium-II simulations trace $2160^3$ ($\sim10$ billion) particles from $z = 127$ to the present day. The Millennium was carried out in a box of original side $500\,h^{-1} \mathrm{Mpc} = 685 \, \mathrm{Mpc}$. After rescaling to the Planck cosmology, the box size becomes $714\,\mathrm{Mpc}$, implying a particle mass of $1.43\times10^9 \Msun$.

From the z = 0 snapshot of the output, we select galaxies with masses above $10^{9.25} \Msun$, for the sake of comparison with the data. This cut results in a sample of 3,369,062 galaxies, i.e. about 240 times larger than the SDSS sample described above. We convert the full 6-dimensional position-velocity data into 3-dimensional observed coordinates ($x$, $y$, redshift) by converting the position and velocity along one Cartesian direction into redshift, omitting the velocities along the other two directions. 

The sSFR distribution of galaxies is unfortunately not identical to the observed distribution, having a long tail towards low sSFR values, and with many galaxies having exactly zero sSFR . In order to try to match the mock data with observations, the galaxies with $\log(\mathrm{sSFR}\;\rm{yr^{-1}})\le-12$ have been assigned a random Gaussian value centered at $\log(\mathrm{sSFR})=-0.3\log(M_*)-8.6$ and with dispersion 0.5 dex. The consequence of this adjustment will be discussed further in Section~\ref{sec:mock_comparison}.

\section{Methodological aspects of the K13 analysis}
\label{sec:met_k13}
In this Section we will examine a number of different aspects of the K13 analysis, highlighting those that are likely to have a significant and deleterious effect on the results. In the following Section~\ref{sec:re_analysis}, we will then modify the analysis to produce new versions of Fig.~\ref{fig:confor_kauff} and show that the long-range conformity signal is likely to be much smaller than indicated in K13, or even absent altogether within the statistical uncertainties.

\subsection{Biases due to density-weighting}
\label{subsec:weight_bias}
\begin{figure*}
\centering
\includegraphics[width=17.9cm]{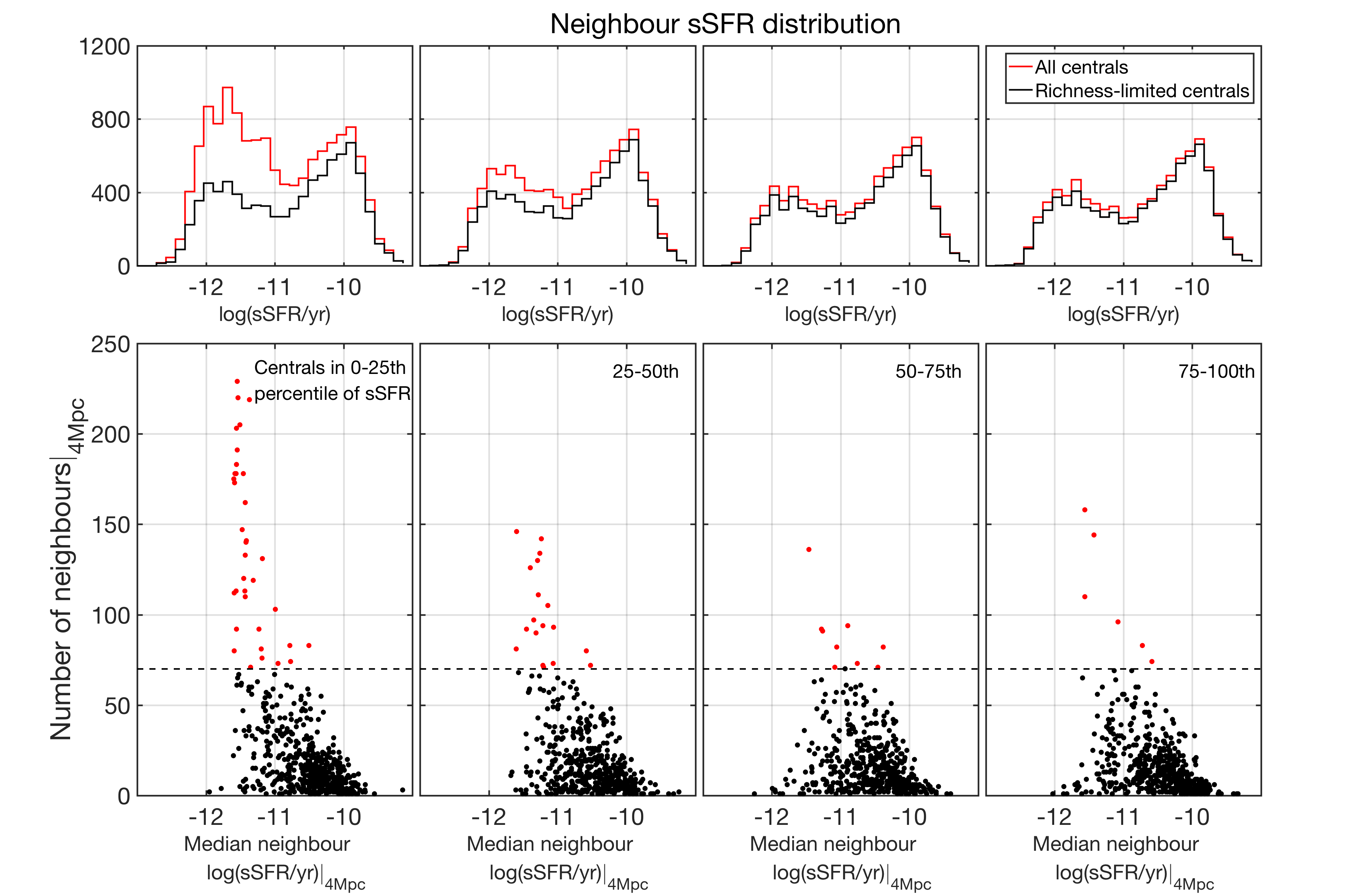}
\caption{The effects of central richness on the neighbour sSFR distribution.
(Top, red) sSFR distribution of neighbours, divided according to the sSFR of their centrals. The panels, from left to right, show the results for centrals in the lowest to highest quartile of sSFR. The striking low-sSFR peak in the leftmost panel corresponds to the significant depression of the red line in Fig.~\ref{fig:confor_kauff}.
(Top, black) Same as red, but only for centrals with 70 neighbours or fewer. The threshold of 70 was chosen arbitrarily to highlight the properties of neighbours of particularly rich centrals.
(Bottom) For centrals in each sSFR quartile, their number of neighbours within 4 Mpc plotted against the median log(sSFR) of their neighbours. Centrals with more than 70 neighbours are highlighted in red. The richest centrals tend to have low sSFRs, and their neighbours tend also to have low sSFRs.}
\label{fig:ssfr_hist}
\end{figure*}

In investigating conformity, we are trying to understand the physical drivers of galaxy evolution, using the star-formation state, e.g. the sSFR, as a probe of the physical conditions locally around each galaxy. If these local physical conditions are somehow correlated over very large scales, then we would see a correlation between the star-formation states of galaxies on similarly large scales.

Within the scale of a single halo, it makes sense to correlate the sSFR of the central with those of the satellites. When investigating conformity on larger scales, the problem is no longer confined exclusively to the satellites of a given central; the general \lq neighbours\rq\ of a central can be satellites in the same halo, or other centrals, or satellites in other nearby haloes. Not least, in K13, the neighbours can have much higher, as well as much lower, stellar masses than the central in question. However, in what follows, we will adhere to the practice in K13, and consider the correlation between the sSFRs of centrals and neighbours.

What should then be done to test the large-scale conformity hypothesis, i.e. that there is a correlation between the sSFRs of centrals and neighbour galaxies on scales extending well beyond individual haloes? Two possible approaches would be as follows:

\begin{enumerate}
\item{Regard each central-neighbour pair as an independent test of the hypothesis that the sSFRs of centrals and neighbours are correlated. A sample of neighbours is constructed by collecting galaxies at a given distance around each and every central in a given sSFR bin. It is then tested to see whether the resulting sSFR distribution of the neighbour sample varies with the central sSFR and/or with the distance-to-central. This was the approach taken in K13.}
\item{Alternatively, one could evaluate the average neighbour sSFR-distance relation for each central, and then average these over all centrals in a given sSFR bin to examine how this relation varies with central sSFR. This approach would be hard to calculate given the discrete nature of galaxies. For instance, for many centrals there would be no neighbour within a given projected distance bin.}
\end{enumerate}

Although the difference between the two example methods may appear superficial, they differ significantly in the weighting of each central in the sample, which may cause differences in the measured conformity signal \citep{Bray2016}. The first approach above implicitly assigns equal weight to every central-neighbour pair, and thereby weights each central by the number of neighbours associated with it (i.e. the \lq richness\rq\ of the environment). In effect, this method preferentially represents the physical processes around \lq rich centrals\rq. In the second approach, all the centrals are weighted equally.

For large-scale conformity (at least in the form which we have stated), we wish to learn about the physical processes reflected by the state of the centrals\rq\ star-formation, and not necessarily by their richness. So, while a physical effect which is probed by many neighbours may be seen with more statistical confidence than one which is probed by just a few, it should not, in our view, be treated with more weight. Therefore, we would favour a method which attaches equal weight to every central, rather than one which weights each central by its number of neighbours (as done in K13).

While analysis approach (ii) avoids weighting centrals by the number of neighbours, it would be difficult to implement in practice. An alternative, which we will implement in this paper (see Section~\ref{sec:re_analysis} below), is to employ approach (i), but to down-weight each central-neighbour pair by the number of neighbours $N_{\mathrm{neigh}}$ that that central has, i.e. that are found within 4 Mpc of the central. In effect, this treats neighbours as probes of the physical processes around centrals, and treats the effects around every central with equal importance, regardless of how many neighbours are influenced by them.

As shown in the bottom panels of Fig.~\ref{fig:ssfr_hist} (the details of which will be discussed shortly), the centrals in the sample display a huge range in the number of neighbours, $N_{\mathrm{neigh}}$. While most have just a few neighbours (the modal value of $N_{\mathrm{neigh}}$ is 2), centrals in the richest environments can have up to $\sim230$ neighbours, and are therefore vastly over-weighted relative to most other centrals. 

Correspondingly, a galaxy in a rich region can be a neighbour to up to 14 centrals, further weighting the effects of dense regions of the Universe in the analysis, and highlighting the difficulty of interpreting conformity on 4 Mpc scales in terms of the effect of a single central galaxy. 

The maximum number of neighbours (as defined here) that are associated with the richest centrals ($N_{\mathrm{neigh}}\sim230$) is strikingly high, given that the centrals under consideration are expected to inhabit relatively low-mass ($\sim10^{11.5}-10^{12}\Msun$) dark matter haloes. The maximum number of 14 centrals per neighbour also indicates a remarkably high density of centrals within the roughly (9 Mpc)$^3$ cylindrical volume for these high-density environments.

The over-weighting of rich environments in K13 is of particular concern given that we would expect (a) that galaxies in high-density regions will preferentially be passive, and (b) any difficulties in isolating true centrals will likely also be more severe in rich environments. We therefore turn to examine these two questions.

In the lower panels of Fig.~\ref{fig:ssfr_hist}, we plot the $N_{\mathrm{neigh}}$ (calculated out to 4 Mpc) and the median neighbour sSFR for each central, splitting the centrals into the four quartiles of central sSFR. The horizontal dashed lines separate centrals with $N_{\mathrm{neigh}}>70$ (plotted in red) from those in less rich environments (plotted in black). The histograms at the top of Fig.~\ref{fig:ssfr_hist} then show the distribution of the sSFRs of the neighbours, again differentiating between the neighbours of centrals with $N_{\mathrm{neigh}}\le70$ (shown as the black histograms) and the whole sample, including the richest environments (red histograms). The difference between the red and black histograms therefore isolates the sSFR distribution of the neighbours of centrals with $N_{\mathrm{neigh}}> 70$ (plotted as the red points in the lower panels).

Fig.~\ref{fig:ssfr_hist} illustrates several important points. First, it shows that the neighbours of low-sSFR centrals tend themselves to have low-sSFR: the leftmost red histogram peaks at much lower sSFR than the other red histograms. This is the conformity signal seen in Fig.~\ref{fig:confor_kauff}. However, it is clear that this is driven by the neighbours of the centrals in the highest density regions ($N_{\mathrm{neigh}}> 70$), rather than by the existence of a strong correlation between a typical central and its 4 Mpc neighbours. Because of the over-weighting of rich centrals, the small number of red points in the lower panels have a disproportionate effect on the histograms in the upper panels. If we ignore all the neighbours of rich centrals (noting that only $\sim$ 3 per cent of centrals in this mass range have more than 70 neighbours), then the neighbour sSFR distributions (shown by the black histograms) are remarkably insensitive to the sSFR of the central, i.e. from left to right in the Figure. It is clear that the rich centrals with $N_{\mathrm{neigh}}> 70$ are themselves concentrated in the first column of the Figure, and are therefore primarily low-sSFR centrals. Because the neighbours in these high-density regions are also of lower sSFR than typical neighbours, it is these high-density neighbours that are almost entirely responsible for the low-sSFR peak in the $0-25^{\mathrm{th}}$ percentile quartile, and therefore for the strong conformity signal. As shown in Fig.~\ref{fig:ssfr_hist} (and as we will show also in Fig.~\ref{fig:sdss_panels}), this small fraction of centrals indeed drives most of the strong, large-scale correlation in Fig.~\ref{fig:confor_kauff}. In the remainder of this Section, we explore the origin of this large-scale correlation, and how a small fraction of centrals can come to produce a dominant effect on the overall neighbour distribution.
\begin{figure*}
\centering
\includegraphics[width=17.9cm]{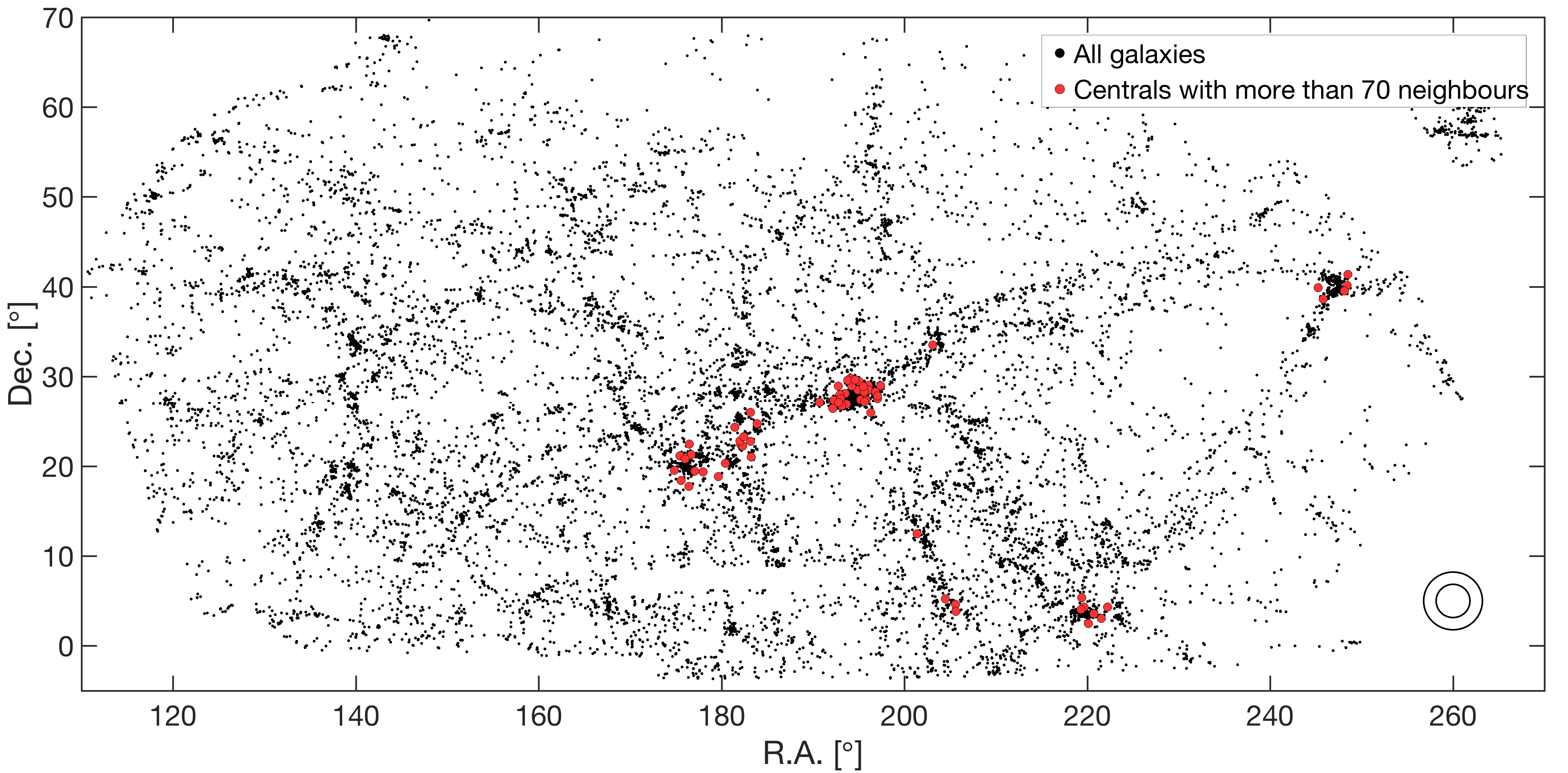}
\caption{The main portion of the SDSS sky, with centrals which have more than 70 neighbours highlighted in red. They are found densely and predominantly near extremely large clusters. The circles at the bottom-right have 4 Mpc radii (the aperture within which a galaxy is considered to be a neighbour) evaluated at the nearest and furthest redshifts of the sample.}
\label{fig:ra_dec_map}
\end{figure*}

Despite the fact that most of the apparent correlation is only driven by a small number of rich centrals, this could nevertheless potentially be an interesting result. These centrals tend to have low sSFRs, and out to a few Mpc, they also tend to have low-sSFR neighbours. This large correlation scale extends well beyond the virial radii of even the most massive dark matter haloes, let alone those which host $10^{10}-10^{10.5} \Msun$ centrals. At face value, this is indeed suggestive of the presence of physical processes operating well beyond the scale of individual haloes.

However, it is instructive to examine where these high-density environments actually occur in the Universe. Fig.~\ref{fig:ra_dec_map} shows that most of the rich centrals (highlighted in red on the Figure) are strongly clustered around just a handful of the very largest galaxy clusters that are present in the SDSS sample. They are not the centrals at the centres of these structures, but instead cluster on their outskirts.

This fact already changes the significance of the 4 Mpc scales for a conformity signal. In the regions of these very largest haloes, environmental-quenching effects will be expected to lower the sSFRs of satellite galaxies over very large regions. For instance, the Coma cluster at [R.A., Dec.] = [195$^{\circ}$, 28$^{\circ}$] has a virial radius of about 2 Mpc \citep{Kubo2007}. Even with a simplistic assumption that satellite quenching extends only out to the virial radius, any galaxy (whether a true central or a satellite of Coma) that is located at the virial radius of Coma would see the passive satellites of Coma extending out to 4 Mpc away, i.e. to a \lq virial diameter\rq\ away. The large scale of the sSFR correlation between this set of galaxies does not therefore correspond to greatly super-halo scales, but rather to the span of the haloes of these extremely large clusters. By introducing an artificial conformity signal to their halo model, \citet{Paranjape2015} have also cautioned that physical effects within single large haloes can easily produce a conformity-like correlation on several Mpc.

Fig.~\ref{fig:ra_dec_map} also emphasizes the very small volume of the Universe that contains these richest centrals. We have argued above that the K13 methodology biases the conformity signal by linearly weighting centrals by the richness of their environments. If we considered a volume-averaging approach to conformity, which could be justified given the use of centrals to probe physical conditions, then we would have a further bias towards rich environments. i.e. an \lq $N^2$\rq\ bias. In Section~\ref{sec:re_analysis}, we will show the dramatic effect of excising very small volumes from the sample.

\subsection{Purity of the central selection}
\label{subsec:purity}
Central galaxies play an important role in conformity, as they are expected to reflect the physical processes near the centre of the dark matter halo. However, the dark matter distribution and potential is difficult to determine in practice, so selecting a complete and pure sample of centrals from an observational catalogue is not a trivial task. As described in Section~\ref{sec:input_data}, K13 used an isolation criterion to identify centrals, i.e. all other galaxies within 500 projected kpc should have less than a half the stellar mass of the central.

It should be noted that misidentification of centrals and satellites would not introduce fake conformity signals if all haloes have the same quenched fraction in the satellites. However, if there is a variation of $\epsilon_{\mathrm{sat}}$ across the sample, i.e. a variation in the quenched fraction of the satellites, then misidentification of the central will more likely lead to a false quenched central in those groups in which the quenched fraction is high, and therefore to a spurious conformity signal.
\begin{figure}
\centering
\includegraphics[width=8.6cm]{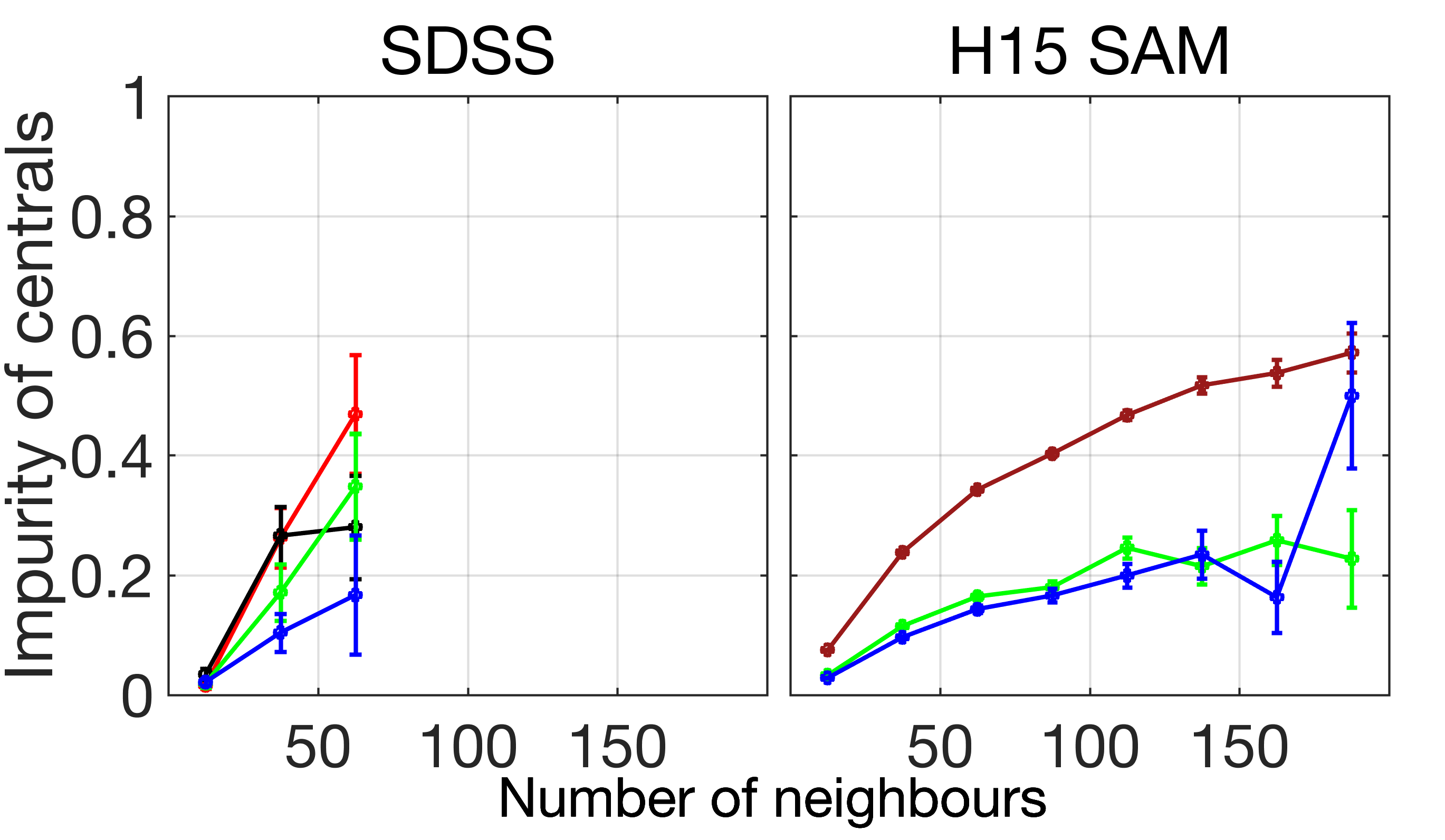}
\caption{ The impurity of the K13 central selection method as a function of number of neighbours. The impurity is defined as the fraction of incorrectly selected centrals (i.e. true satellites) out of all selected centrals.
(Left) The impurity of the K13 central selection in SDSS, as a function of number of neighbours within 4 Mpc. We use the Yang et al. group catalogue as a reference for \lq true\rq\ centrals using the central identification algorithm described in Section~\ref{subsec:group_cat}. The red, black, green, and blue symbols respectively indicate centrals in the $0-25^{\mathrm{th}}$, $25-50^{\mathrm{th}}$, $50-75^{\mathrm{th}}$, and $75-100^{\mathrm{th}}$ percentile of sSFR.  To avoid displaying spuriously high or low impurity fractions, bins containing $0 < N \le 10$ points are omitted.
(Right) The impurity of the K13 central selection in the H15 SAM. Since the sSFRs of low-sSFR galaxies are scrambled, we treat centrals in the lowest two quartiles of sSFR as a single set, and represent their neighbours as a single brown line (see Section~\ref{subsec:mock_cat}, and the detailed discussion in Section~\ref{sec:mock_comparison}).}
\label{fig:fig_purity}
\end{figure}
\begin{figure*}
\centering
\includegraphics[width=17.9cm]{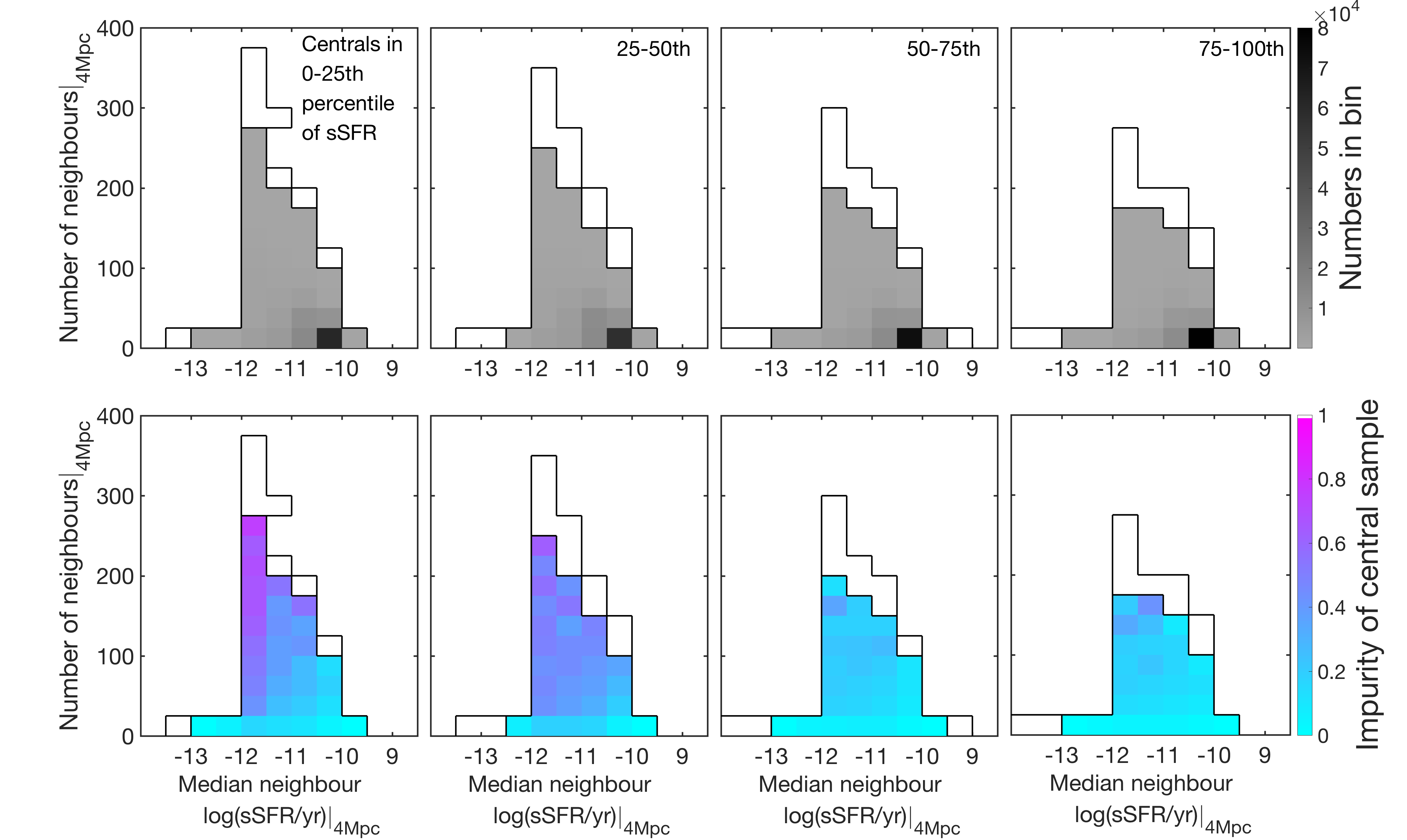}
\caption{Impurity of the K13 central selection method when applied to the H15 SAM, as a function of number of neighbours and neighbour sSFR. 
(Top) Number distribution of centrals as a function of number of neighbours and median log(sSFR) of neighbours.
(Bottom) Central impurity as a function of number of neighbours and median log(sSFR) of neighbours. To avoid displaying spuriously high or low impurity fractions, grid elements with $0 < N \le 10$ points are blanked out and outlined in black. In high-density, low-sSFR environments, up to 2 out of 3 selected centrals are satellite contaminants.}
\label{fig:2d_purity}
\end{figure*}

An alternative to using an isolation criterion is to try to identify groups in the galaxy catalogue. A number of group-finding algorithms exist for this purpose. Assuming that these groups really trace the dark matter distribution, one can then try to identify the galaxies at the minima of the potential wells, i.e. the true centrals.

It is then slightly worrying to find in Fig.~\ref{fig:ra_dec_map} that many of the centrals identified by the isolation criterion are found in and around large clusters, especially considering that the true centrals of these clusters are expected to be significantly more massive than the $10^{10}\Msun < M_* < 10^{10.5}\Msun$ range under consideration. The selection criterion of comparing the masses of galaxies with their near (500 kpc)  neighbours allows satellites in large clusters to be identified as centrals, as long as they are sufficiently more massive than their near neighbours. 

K13 justified the use of this selection method by applying it to mock catalogues from the semi-analytic model of \citet{Guo2011}. The contamination of the central sample from satellites varies as a function of mass, but was at most 30 per cent. However, given the existence of the biasing towards high-density regions discussed in the previous sub-section, even a few satellite contaminants from large clusters could result in a disproportionately large contamination in the results. 

We therefore re-examined the effectiveness of the K13 central selection method in the context of environment richness and sSFR. In order to do this, we compare the isolation criterion of K13 with central-satellite classifications made with reference both to the Yang et al. SDSS group catalogue as well as to the H15 semi-analytic model. For the latter, the true central-satellite identities of all galaxies are of course known. In both cases we compute the impurity, defined as the fraction of satellites present in the isolation-selected sample of centrals. Fig.~\ref{fig:fig_purity} shows this impurity as a function of $N_{\mathrm{neigh}}$ for centrals in the four quartiles of central sSFR.

Fig.~\ref{fig:2d_purity} shows the number and impurity of the isolation-selected centrals in the H15 SAM as a bivariate function of $N_{\mathrm{neigh}}$ and the median log(sSFR) of the neighbours, restricting attention to the SAM to exploit the much larger number of groups for the bivariate analysis. We find that for identified centrals within $10^{10}-10^{10.5}\Msun$, the global contamination fraction is indeed low (we find about 10 per cent). However, in regions of high number density (i.e. near large clusters), which also generally have low sSFR, we find that up to two-thirds of selected centrals are in fact satellites (see Fig.~\ref{fig:fig_purity} and \ref{fig:2d_purity}). Given that the volume of the isolation criterion (500 kpc radius, $\pm \,500\, \mathrm{km}\,\mathrm{s}^{-1}$) is relatively small compared to the dimensions of the largest clusters (of order 2 Mpc radius, and  $\sigma_\mathrm{v} \sim1000\, \mathrm{km}\,\mathrm{s}^{-1}$), it is perhaps unsurprising that, when applied on the outskirts of these clusters, the isolation criterion is misclassifying satellite galaxies as centrals. As previously illustrated, these clusters have high number density, and have environmentally-driven quenching that extends over several Mpc. These are therefore precisely the regions where the density-bias discussed in the previous sub-section will have the greatest effect on the results. Any small mistake in identifying centrals and satellites will produce a spurious conformity signal (since the galaxies will have low sSFRs) which will then be strongly amplified by the density-weighting to have a disproportionate effect on the final result.

We conclude that out of the very small fraction of centrals (3 per cent) with $N_{\mathrm{neigh}}> 70$, which are likely to be responsible for most of the observed conformity signal, more than half are probably satellite contaminants. It is however worth noting that it is likely that not all of these problematic rich centrals are contaminants. Some are likely to be real centrals at the centres of their respective haloes, but which reside just beyond the extent of much larger clusters. The key point is that their numerous neighbours then correspond to the members of these large clusters. If they are indeed true centrals, then the sSFR correlation seen between them and the members of these large clusters would be indicative of a real conformity effect, albeit one affecting only a very small fraction of centrals. The implications of this will be discussed further in Section.~\ref{sec:re_analysis}.

\subsection{Statistical summaries of the sSFR distributions}
\label{subsec:stat_sum_ssfr}

Both the splitting of the centrals and the analysis of neighbour properties were based, in K13, on the sSFR  distributions of the centrals and neighbours respectively. K13 split centrals into quartiles, and summarized the sSFR of neighbours by using the median. 

While a normal distribution is completely characterized by its mean and variance, the distribution of sSFRs is clearly not Gaussian, and the consequences of the choice of summary statistic are worth considering. The median, as used by K13, has a number of attractive features. In a unimodal distribution, the median has the benefit of being insensitive to outliers in the wings of the distribution. This is especially important in the case of passive galaxies, for which the estimation of individual sSFR has significantly higher uncertainty than for Main Sequence galaxies, with uncertainties of order 1 dex \citep{Brinchmann2004}. The effects of these uncertainties in the individual sSFR are relatively limited with the use of the median.

However, the sSFR distribution of galaxies is known to be bimodal, because of the effects of quenching. This is clearly seen in the histograms of neighbour sSFR in Fig.~\ref{fig:ssfr_hist}, where the two modes (corresponding to the star-forming and the passive population) are of comparable strength. If one mode is dominant and has a small dispersion, a small change in the relative numbers of galaxies in each component will have little effect on the median. But, if both components are narrow and of equal strength, then a small shift in their relative size can produce a very large change in the median, as the median jumps from one mode to the other. 

\begin{figure}
\centering
\includegraphics[width=8.6cm]{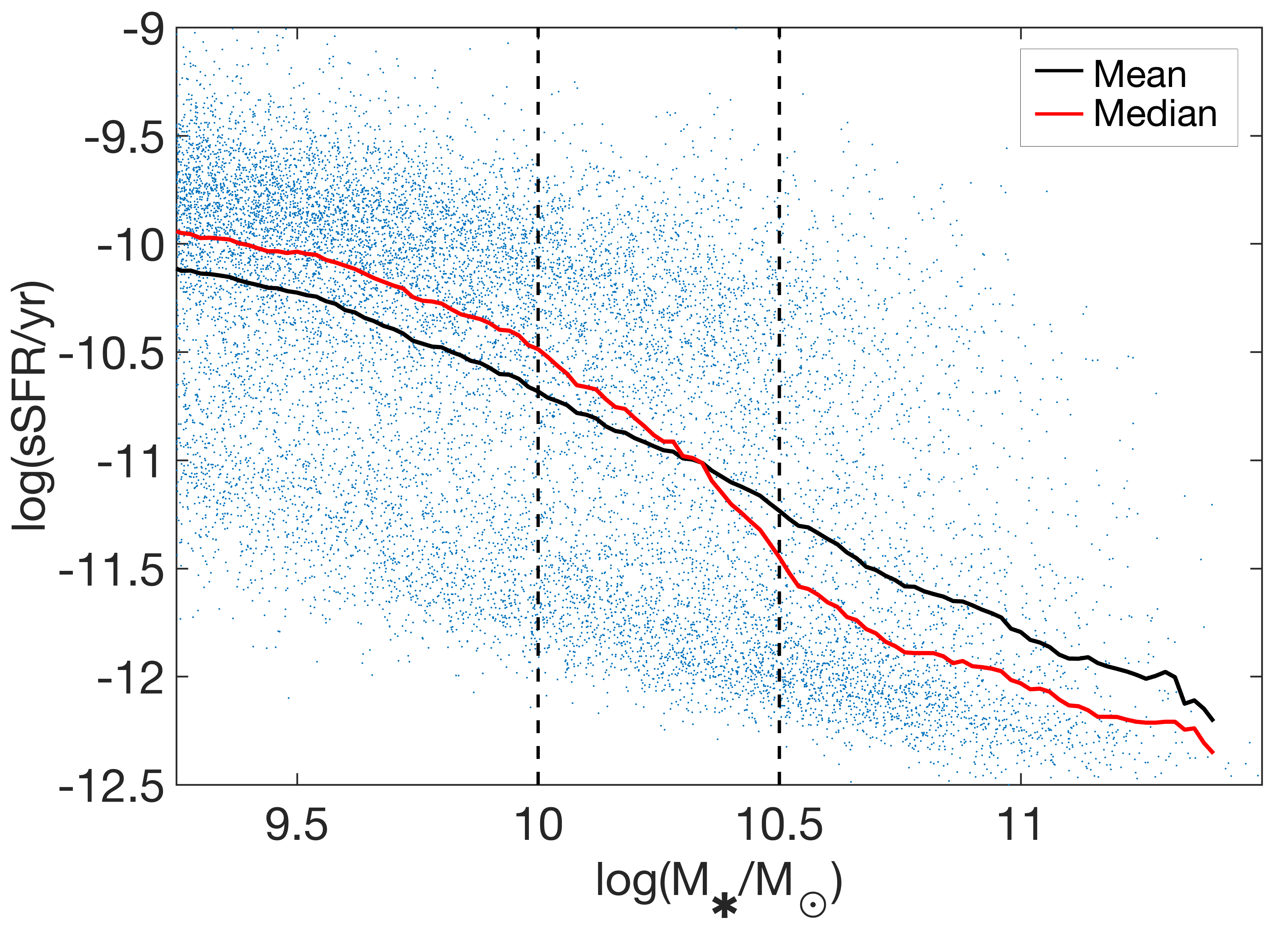}
\caption{Distribution of the data as a function of sSFR and stellar mass. The solid black and red lines represent respectively the mean and median log(sSFR), evaluated with running bins of 0.2 dex width in stellar mass, and serve as a comparison of how the two different statistics behave under a bimodal distribution with an abruptly changing major mode. The two vertical lines mark the range of the centrals under consideration.}
\label{fig:running_averages}
\end{figure}

This is illustrated in Fig.~\ref{fig:running_averages}, where we simply plot the sSFR of the galaxies in our complete sample as a function of their stellar mass. The black and red curves respectively show the mean and median log(sSFR) calculated in a running bin of width 0.2 dex in mass. The running mean varies smoothly with mass. The running median, however, varies much more slowly at high and low masses, where the distribution is dominated by one or other of the modes, but varies much faster than the mean in the region where the distribution is evenly divided between the two components. In essence, the choice of median to characterize a strongly bimodal distribution with roughly equal components has the effect of amplifying changes relative to what would be seen in the mean. 

In the specific case considered here, it is clear from Fig.~\ref{fig:ssfr_hist} that we are in the regime where both peaks of the neighbour sSFR distribution are of comparable strength. Indeed, it is clear in the top panels of Fig.~\ref{fig:ssfr_hist} that the methodological bias in favour of the highest density centrals has tipped the balance between the two modes. The low-sSFR component goes from being slightly sub-dominant (in the three right-hand histograms) to slightly dominant (in the leftmost panel in Fig.~\ref{fig:ssfr_hist}). The choice of median to characterize the sSFR distribution therefore unfortunately further amplifies the effect of the density-bias.

A further possible effect arises from the splitting of the centrals into quartiles of sSFR within their mass range $10^{10}\Msun < M_* < 10^{10.5}\Msun$. While the sSFR-mass correlation in both the star-forming and passive modes is not a strong function of mass, the relative strength of the two modes changes with mass as a result of the mass-dependence of mass-quenching. As shown in Fig.~\ref{fig:running_averages}, the relative size of the two modes shifts dramatically in precisely the galaxy mass range where (for the centrals) the K13 conformity signal is strongest. This means that the mass distribution, within the $10^{10}\Msun < M_* < 10^{10.5}\Msun$ range, of the centrals in the different sSFR quartiles will be different, and therefore quite possibly the host halo mass distributions. This has two implications. Ideally, a much more stringent matching of the masses of the centrals would be required to remove the possibility of conformity-like signals arising from straight-forward correlations with halo mass (see \citealt{Knobel2015} for further discussion). 

An additional consequence is that, within the 500 kpc radius of the isolation criterion for the identification of centrals, the mass distribution of the neighbours will be different from the general neighbour population, and furthermore different for the different quartiles of central sSFR. This causes the pronounced upturn in the sSFR of neighbours within 500 kpc in Fig.~\ref{fig:confor_kauff}. For this reason, the region within 500 kpc must be clearly differentiated from the larger scales, as indicated on Fig.~\ref{fig:confor_kauff} and later Figures in the paper.

A further point is that characterizing the star formation activity of the centrals using sSFR-quartile bins means that centrals in the same quartile (in particular, the $25-50^{\mathrm{th}}$ and $50-75^{\mathrm{th}}$ percentiles) may be somewhat heterogeneous in their star formation state, i.e. whether they are star-forming or passive. That is to say, star-forming and passive galaxies will be classified as having similar \lq star-formation activity\rq\ because they lie in the same sSFR-quartile. At high masses (above $10^{11} \Msun$), the problem is reversed, in the sense that all centrals are passive, and so the different quartiles in the sSFR distribution contain a rather homogeneous set of passive centrals. This is likely the cause of the reduction of the conformity signal for these more massive centrals in fig. 3 of K13.

\subsection{Summary of the methodological aspects of K13}
\label{subsec:k13_sum}
To summarize the previous discussion, we have identified at least three aspects of the methodology adopted in K13 which may have produced biases or amplifications of conformity-like signals in their analysis, thereby producing misleading results.

\begin{enumerate}
\item{Bias due to density-weighting:  By giving equal weight to every central-neighbour pair, the methodology drastically over-represents central galaxies in high-density regions, i.e. those in the neighbourhoods of the largest clusters. In doing so, it allows conformity signals to appear on the spatial scales of these largest clusters, rather than on the scales of the relatively low-mass haloes that are nominally being probed with these quite low-mass centrals.} 

\item{Central selection: The isolation criterion, despite making no reference to group catalogues, performs quite well overall in identifying central galaxies. However, the contamination from satellites increases markedly in high-density regions, where the fraction of passive satellites is also high. This can produce a spurious conformity signal that is then amplified by the density-weighting discussed above. }

\item{Representation of sSFR distributions: While the choice of using the median to represent the distribution of neighbour sSFRs has some benefits, it has the unfortunate result of amplifying the apparent strength of the conformity signal, whether real or produced by the above effects, when the two bimodal components are of comparable amplitude. The choice of using sSFR percentiles to represent the star-formation activity of the centrals also groups together heterogeneous subsets of centrals, and produces a clear bias in the region within the 500 kpc radius used in the isolation criterion for selecting central galaxies.}
\end{enumerate}

In the following Section, we demonstrate that the combination of these different methodological aspects grossly amplifies the sSFR correlations that are present in a very specific subset of the data, and may therefore produce misleading results in the minds of most readers.
\begin{figure*}
\centering
\includegraphics[width=17.9cm]{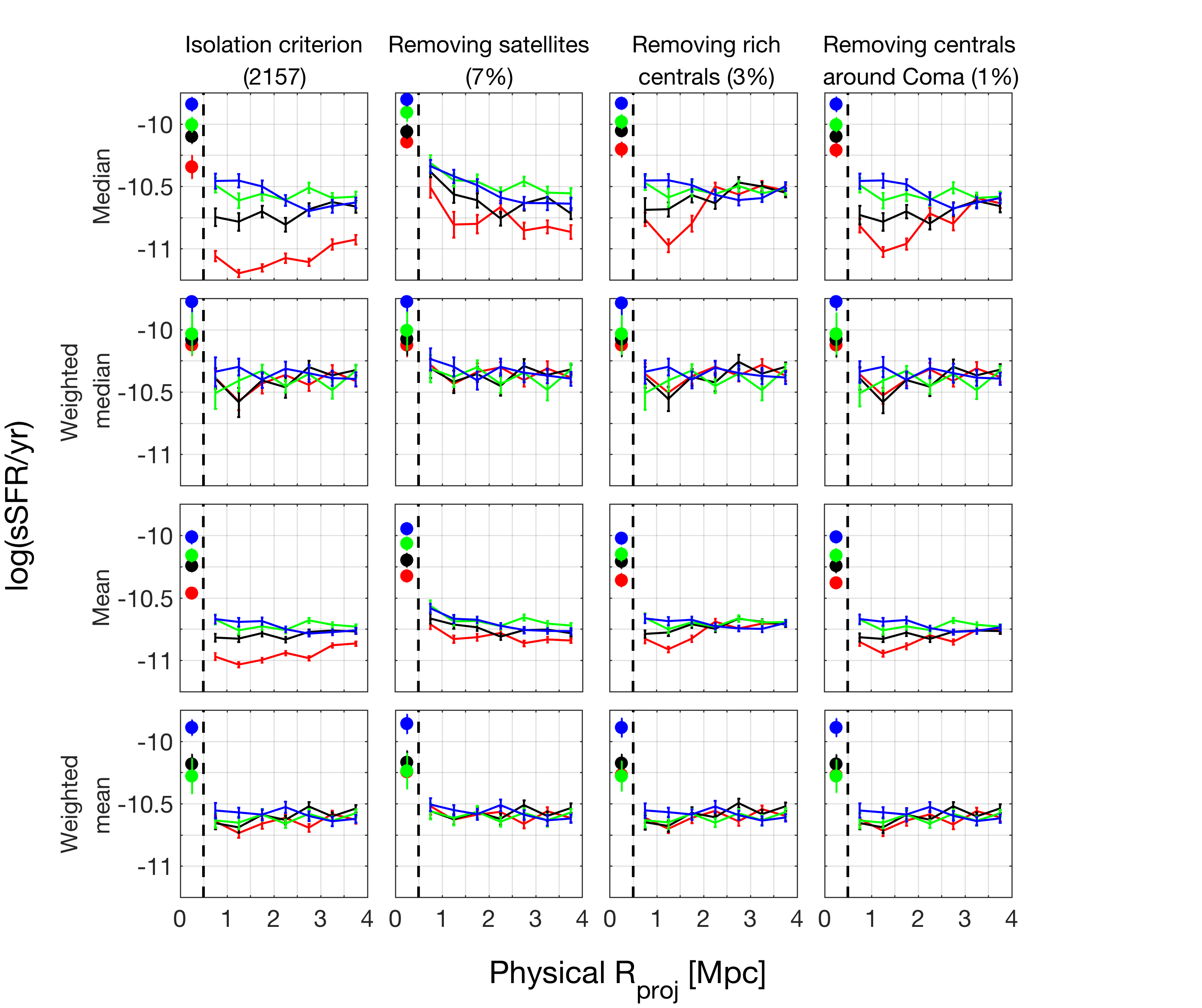}
\caption{Variants of Fig.~\ref{fig:confor_kauff} that result from modifying the methodology of K13. The columns show the results of the different cuts made to the central sample, and the rows show the results of the different summary statistics used. The first row and the first column are the choices used in the original K13 analysis, so the top-left panel is the same as Fig.~\ref{fig:confor_kauff}. The percentage quoted in each column title refers to the fraction of centrals that are removed in that particular cut. The $25^{\mathrm{th}}$ and $75^{\mathrm{th}}$ percentiles of the sSFR distribution are omitted from the plots for clarity. Note that as a result, there is a difference in the $y$-axis limits between this Figure and Fig.~\ref{fig:confor_kauff}.}
\label{fig:sdss_panels}
\end{figure*}
\begin{figure*}
\centering
\includegraphics[width=17.9cm]{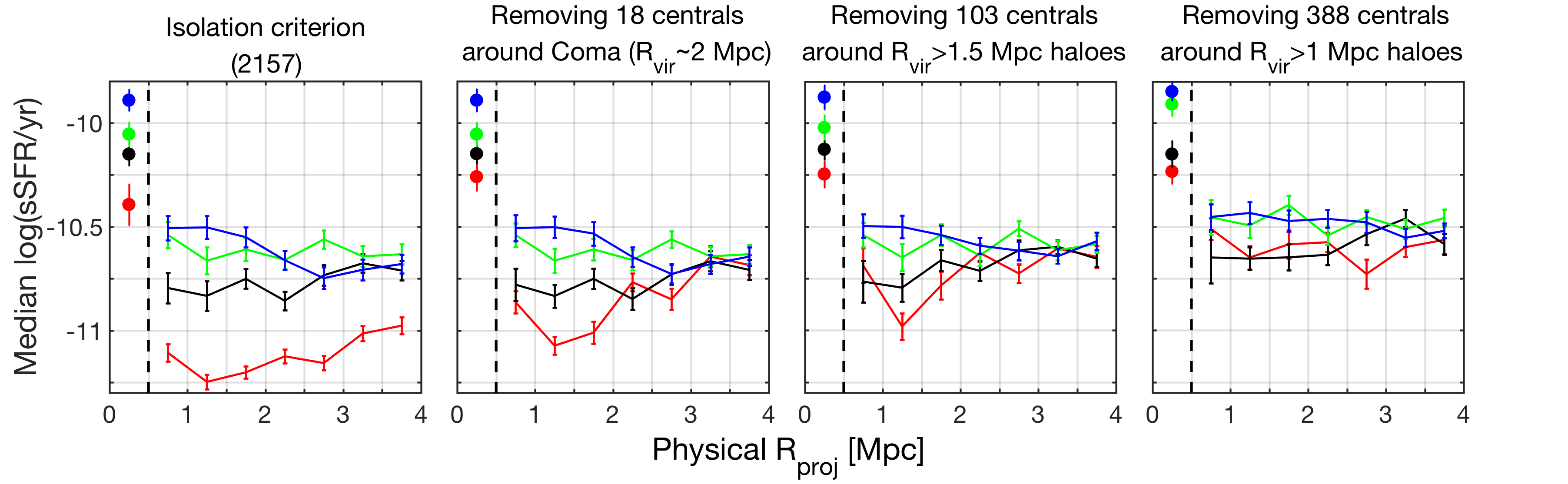}
\caption{Variants of Fig.~\ref{fig:confor_kauff} that result from removing centrals around progressively smaller haloes. The leftmost panel is analogous to Fig.~\ref{fig:confor_kauff}. The other panels show, respectively, variants of Fig.~\ref{fig:confor_kauff} when removing centrals around haloes with $R_{\rm{vir}}>$ 2, 1.5, and 1 Mpc. Centrals are cut based on the same $R_{\mathrm{proj}} = 4 \, \mathrm{Mpc}$ and $c\Delta z = \pm \,1000 \, \mathrm{km\,s^{-1}}$ cylinder criterion described in operation (ii) above. The virial radii are derived from the virial mass estimates in the Yang et al. group catalogue. The quoted numbers in the title of each panel refer to the cumulative number of removed centrals at each step.}
\label{fig:rem_cyl}
\end{figure*}

\section{Re-analysis under modified methodology}
\label{sec:re_analysis}
In this Section, we explore the degree to which the effects discussed in Section~\ref{sec:met_k13} actually affect the data. In order to do so, we apply simple modifications to the K13 analysis which specifically address these issues, either by adjusting the given methodology, or by making different methodological choices. We emphasize that while these ad hoc modifications do, we believe, give the data a more fair representation, our purpose is to illustrate the compounded effect of the various biases on the K13 results, and not to make a serious attempt to quantify galactic conformity within this framework. We plan this for a later paper. We also note that some recent studies of large-scale conformity also examine the effects of these methodological choices \citep[e.g.][]{Bray2016, Berti2017}. The consistency between their findings and this work will be briefly discussed in Section~\ref{sec:discussion}.

In the following, we discuss possible ways to counteract each of these effects. We then apply them to the data, exploring the extent to which the apparent conformity signal persists as one or more of them are applied.\newline

The fact that centrals are weighted in proportion to their richness results in a strong bias towards large clusters. In addressing this density-weighting bias, we apply the following measures (separately or together).

\begin{enumerate}
\item{\textbf{Remove the richest centrals.} By removing all centrals which have more neighbours than a somewhat arbitrary limit of $N_{\mathrm{neigh}} = 70$, we can remove that subset of centrals which most strongly bias the result. Following from the discussion in Section~\ref{subsec:rep_k13}, this action also removes the subset of centrals for which the contamination fraction from satellites is highest. This cut removes just 68 (3 per cent) of the centrals within the relevant mass range $10^{10}\Msun < M_* < 10^{10.5}\Msun$. We note that about a half of these excised galaxies (31 out of 68) were in fact classified as satellites in the group catalogue.}

\item{\textbf{Remove centrals near to the most massive cluster.} As an alternative to (i), we simply remove centrals which are near to the single largest cluster in the SDSS data, which is the Coma cluster. We do so by excluding all centrals that are located within a cylinder of $R_{\mathrm{proj}} = 4 \, \mathrm{Mpc}$ and $c\Delta z = \pm \,1000 \, \mathrm{km\,s^{-1}}$ that is centred on Coma. The centre of Coma is defined here as the median R.A., dec., and $z$ of the members of Coma, where the group memberships are defined according to the Yang et al. group catalogue. It should be noted that this does not remove the central of Coma itself, as it lies well above the central mass range under consideration ($10^{10} - 10^{10.5} \Msun$). This cut removes only 18 (1 per cent) of the centrals within the relevant mass range, i.e. it excludes much fewer than (i). Of the 18 excluded centrals, 11 were classified as satellites in the group catalogue, suggesting that there are some genuine centrals existing close to these largest clusters. Unsurprisingly, all 18 of these centrals have more than 70 neighbours, i.e. they are all also removed by operation (i).}

\item{\textbf{De-weight rich environments.} Apart from excising centrals on the basis of richness, we can also simply down-weight each central-neighbour pair by the total number of neighbours within 4 Mpc of the central, producing a result in which all centrals are equally weighted (see discussion in Section~\ref{subsec:weight_bias}). In computing median sSFRs with the down-weighted samples, we simply compute the $50^{\rm{th}}$ percentile point in weight. \newline

In addressing the specific issue of impurity of the central sample, i.e. the contamination from galaxies that are actually satellites, the straightforward solution is to:}

\item{\textbf{Remove all suspected satellites from the central sample.} We make use of the Yang et al. group catalogue to identify groups and their centrals in the SDSS data (as was described in Section~\ref{subsec:rep_k13}). We then remove all satellites (i.e. non-centrals) from the sample of \lq centrals\rq\ that was selected by the K13 isolation criterion. Since the group finder has demonstrably good performance on large haloes ($M_{\mathrm{vir}} \gtrsim 10^{12} h^{-1} \Msun$; \citealt{Yang2007}), it is well-suited to identify potential contaminants in high-density regions where, as we have seen, they have the greatest impact on the analysis. A total of 168 galaxies (7 per cent) are removed from the central sample in this way. \newline

To address the issues with the use of the median, we simply:}

\item{\textbf{Use the mean (of the log) instead of the median.} Note that we can make use of both the non-weighted and down-weighted samples to compute these means.} 
\end{enumerate}

In Fig.~\ref{fig:sdss_panels}, we show the effect of applying these various modifications to the K13 analysis, either independently or in combination with each other. The upper left panel reproduces the original K13 analysis from Fig.~\ref{fig:confor_kauff} of this paper. Subsequent rows downwards in Fig.~\ref{fig:sdss_panels} show the effect of down-weighting pairs by $N_{\mathrm{neigh}}$ (i.e operation (iii) above), of computing non-weighted means (i.e. (v) above), and finally of computing down-weighted means (both (iii) and (v) together). The second column in Fig.~\ref{fig:sdss_panels} shows the effect of removing suspected satellites from the central sample (i.e. (iv) above). The next column shows the effect of instead simply removing all centrals with $N_{\mathrm{neigh}}> 70$ (i.e. (i) above), while the rightmost column shows the result of instead removing those 18 centrals lying in a cylinder centered on the Coma cluster (i.e. (ii) above).

It is clear from Fig.~\ref{fig:sdss_panels} that all of the methodological modifications described above have, as one would expect, the effect of decreasing the amplitude and/or spatial scale of the conformity signal. 

Detailed comparisons of the panels in Fig.~\ref{fig:sdss_panels} are also consistent with our previous discussion in Section~\ref{sec:met_k13}. We first discuss the effects of individual methodological modifications on the conformity signal, in comparison with the original K13 methodology. The difference between using the median and the mean is primarily to reduce the amplitude independently of scale, as would be expected. Interestingly, removing the relatively large number (7 per cent) of likely satellite contaminants also mostly affects the amplitude and not the scale. In both cases, the depression of the sSFR of the neighbours of the lowest-sSFR centrals (red line) relative to the others is reduced approximately by a factor of 2 (in the average of the logarithm).

More importantly, Fig.~\ref{fig:sdss_panels} illustrates the disproportionate effect of the density-weighting of centrals on the original K13 result. By weighting each central-neighbour pair by $({N_{\mathrm{neigh}}})^{-1}$ (i.e. the $2^{\rm{nd}}$ and the $4^{\rm{th}}$ row), the conformity signal beyond 1.5 Mpc completely disappears, while the remaining sub-1.5 Mpc signal is substantially weakened. In the case of the weighted median, the depression of the red line relative to the blue line at 1\,Mpc is $\sim$\,0.2 dex, while the same signal in the weighted mean is $\sim$\,0.1 dex. In both cases, the amplitude of the remaining signal is comparable to the bootstrap uncertainties, and is therefore difficult to distinguish from noise. That is to say, when we treat the effects around centrals with equal weight, regardless of their local density, there is already very little evidence for the existence of large-scale conformity.

By taking all of these methodological modifications together simultaneously (i.e. in the $4^{\rm{th}}$ row, $2^{\rm{nd}}$ column of Fig.~\ref{fig:sdss_panels}), we measure a conformity signal which we believe to be more robust and less biased. In order to place an upper limit on this remaining conformity signal, we compare the depression of the red line relative to the blue line in that panel. We find that the signal is at most 0.08\,dex (at 1.25\,Mpc); this is comparable to the respective bootstrap uncertainties at this radius, which are $\sim$\,0.05\,dex. The data therefore does not support the existence of a conformity signal at any scale beyond that of the virial radius. In Section~\ref{sec:mock_comparison}, we further discuss the interpretations of this null result in the context of the limited cosmological volume of this data set.

Out of the three methodological issues that we have addressed, the implicit bias towards high-density regions produces the most dramatic amplification of the conformity signal. In fact, the simple operation of removing the 18 centrals in the 4 Mpc cylinder around the Coma cluster is already enough to essentially remove the large-scale signal beyond 2 Mpc. This emphasizes the fact that the large-scale conformity in the overall SDSS sample that is seen in K13 is mostly associated with the very small number of the very largest haloes, rather than with super-halo scale effects associated with the relatively low-mass haloes that host the nominal set of centrals used in the analysis.

The rightmost column of Fig.~\ref{fig:sdss_panels} serves to highlight the dramatic effect of density-weighting from a few centrals which are around the very largest cluster, and shows that most of the large-scale conformity seen in K13 is in fact driven by the very largest haloes. However, as the removal of density-weighting (i.e. $2^{\rm{nd}}$ and $4^{\rm{th}}$ row of Fig.~\ref{fig:sdss_panels}) demonstrates, the remaining sub-2\,Mpc signal is also driven mostly by density-weighting, presumably from centrals around clusters which are smaller, but more common, than Coma.

Fig.~\ref{fig:rem_cyl} illustrates this point more clearly. The leftmost panel of Fig.~\ref{fig:rem_cyl} is analogous to Fig.~\ref{fig:confor_kauff}, while the other panels, from left to right, show the cumulative effect of removing centrals around relatively large haloes of progressively smaller sizes, from the one system with \lq virial diameter\rq\ of $\sim$ 4 Mpc (Coma), down to those with virial diameters of only 2 Mpc. At each step of the cuts, the original conformity signal at the corresponding virial diameter is completely eliminated, while the signal at shorter ranges is substantially reduced. After removing centrals around systems with $R_{\mathrm{vir}}>1\,\mathrm{Mpc}$, there remains only a very weak conformity signal. Comparing the sSFR of neighbours of relatively low-sSFR centrals (red and black lines) with those of relatively high-sSFR centrals (green and blue lines), one finds a much weaker systematic depression of $\sim$\,0.2\,dex out to separations of $\sim$\,2\,Mpc. This further illustrates the fact that, due to the effects of density-weighting, the large-scale conformity presented in K13 is primarily driven by effects on the virial scales of the largest clusters.

These results still allow that, for a very small number of centrals around the richest clusters, there may be a residual real conformity effect on scales just beyond the virial radius. Such an effect could arise from large-scale spatial correlations of halo accretion rates, as illustrated by \citet{Hearin2016}. Their analysis of dark matter simulations showed that, due to large-scale tidal interactions, haloes residing in high-density environments tend to have lower dark matter accretion rates. Such a spatial correlation spans over several Mpc. Since the SFRs of galaxies and the accretion rates of their host haloes are expected to be positively correlated, this \lq halo accretion conformity\rq\ could be a driver of some degree of large-scale galactic conformity.

Alternatively, it could be indicative of other physical processes acting beyond the halo, such as energetic feedback from active galactic nuclei residing within the larger cluster, which may eject hot gas somewhat beyond the virial radius, and thereby suppress star-formation on super-halo scales (see \citealt{Kauffmann2015}). 

In such cases, the conformity effect, even if real, would clearly have been driven by the larger system, and should not be thought of as having been driven by the relatively low-mass centrals under consideration. The corresponding length scale of the conformity signal should therefore be compared with the size of the larger system, and not with that of the smaller halo.

\begin{figure*}
\centering
\includegraphics[width=17.9cm]{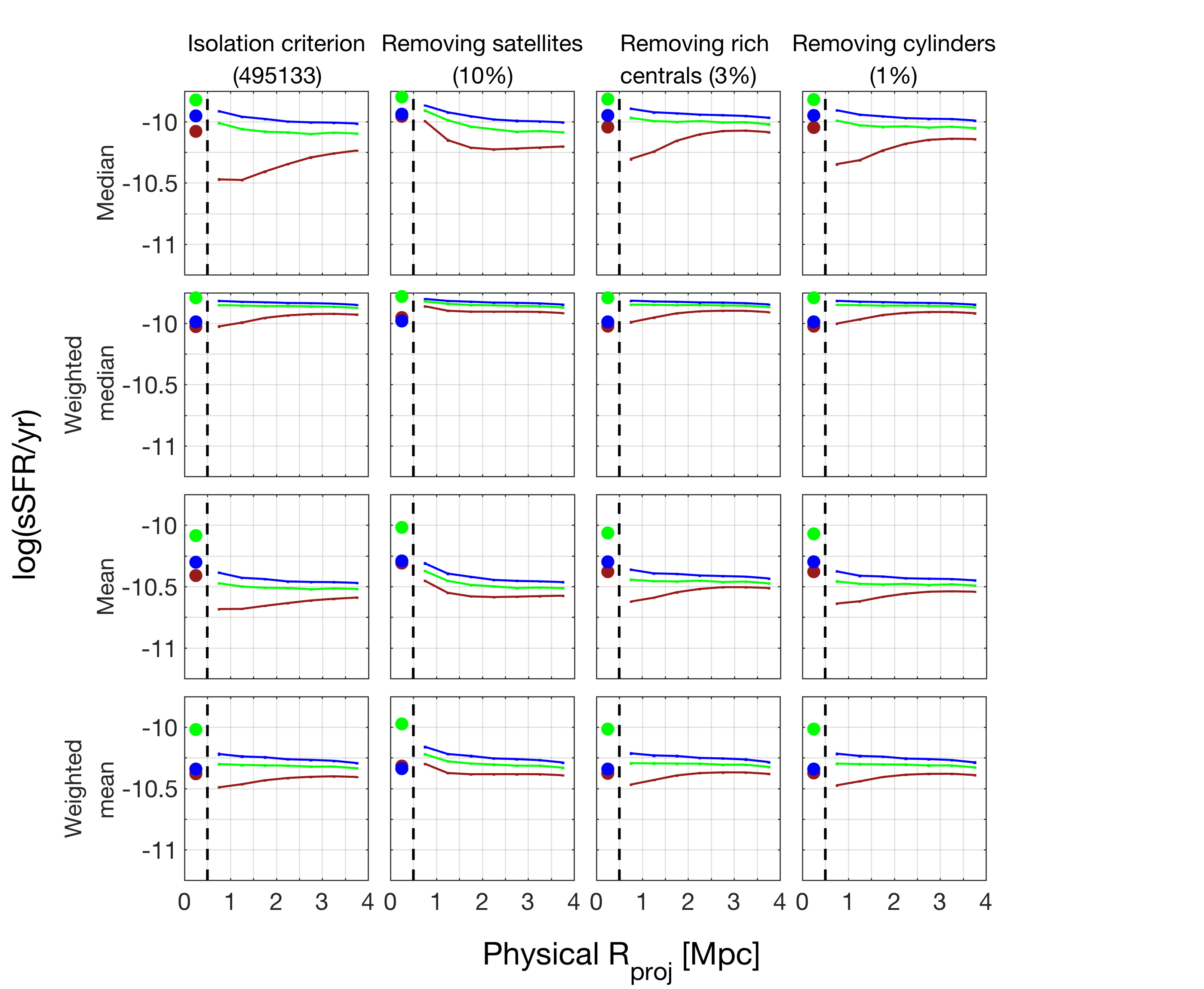}
\caption{Same as Fig.~\ref{fig:sdss_panels}, applied to the H15 SAM. Since we treat centrals in the lowest two quartile as a single sample, we represent their neighbours with a single brown line (see text for details).}
\label{fig:mock_panels}
\end{figure*}
\begin{figure*}
\centering
\includegraphics[width=17.9cm]{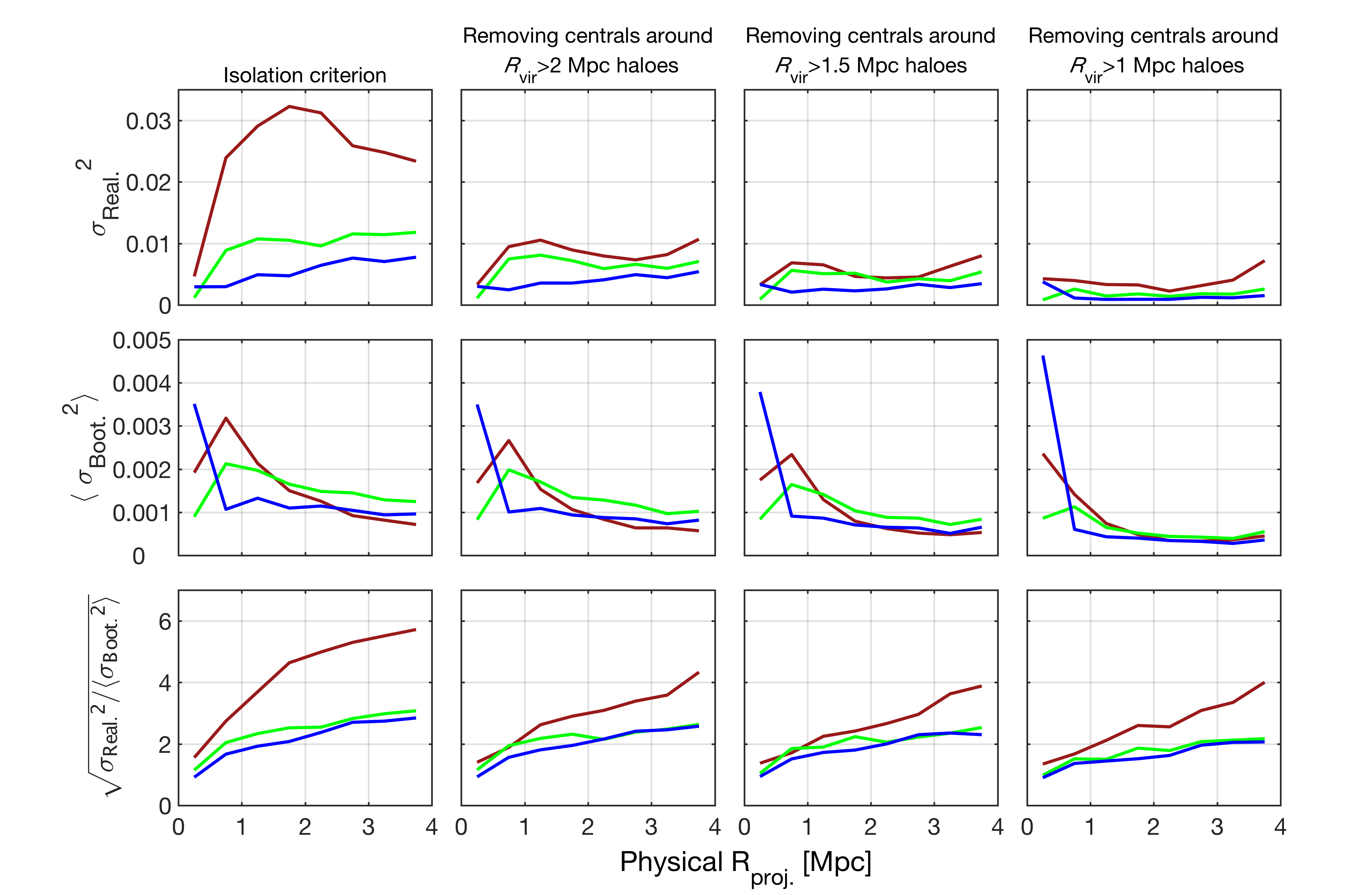}
\caption{The sources of variance in the sSFR distribution of neighbours in SDSS-like volumes in the H15 mock. The neighbours are split by central sSFR and colour-coded in the same way as Fig.~\ref{fig:mock_panels}. From the simulation volume, 125 non-overlapping sub-volumes were selected, each with spatial dimensions of $140\,\mathrm{Mpc}\times 140\,\mathrm{Mpc}$ and $\Delta z = 0.02$.  The columns correspond to cuts to the central sample which are analogous to those made in Fig.~\ref{fig:rem_cyl}. (Top) The variance in the sSFR distribution of neighbours across the 125 realizations. (Middle) The median value of the bootstrap variances, averaged over the 125 realizations. (Bottom) The square root of the ratio of the two above quantities, i.e. the relative importance of cosmic variance and Poisson noise.}
\label{fig:realn_noise}
\end{figure*}

\section{Comparison with the H15 semi-analytic model}
\label{sec:mock_comparison}

Although the existence of conformity is well-established between centrals and satellites of the same halo \citep{Weinmann2006, Knobel2015}, one does also expect some degree of sSFR correlation on the scale of several Mpc as a result of the fact that halo properties will be correlated on large scales. An example is the halo assembly history, or concentration \citep{Hearin2015, Paranjape2015, Hearin2016}.

In order to determine how much correlation is expected from known and predictable physical processes, one can apply an identical analysis to mock catalogues generated from semi-analytic models of galaxy evolution, within which the semi-analytical prescriptions of baryonic physics govern the evolution of simulated galaxies within their respective dark matter haloes. A significant feature in K13 was the fact that, while it was claimed that strong, long-range, conformity-like correlations existed in the data, similar signals were only very weakly present in the parallel analysis of the \citet{Guo2011} SAM. This indicates that there is a physical process (or processes) operating in Nature that is not included in the SAM. This possibility was further discussed in \citet{Kauffmann2015}.

We have demonstrated in this paper that most of the signal is driven by sSFR correlations in high-mass haloes via the density-weighting effect, further amplified by a number of other aspects of the analysis. Even so, prescriptions of environmental quenching are present in SAMs, and so such an effect should be present to a comparable degree if the mock catalogue is analysed in the same way.

In order to explore this, we apply the same suite of methodologies that we described in Section~\ref{sec:re_analysis} to the mock catalogue from the H15 SAM. The results are presented in Fig.~\ref{fig:mock_panels}, the different panels of which are directly analogous to those in Fig.~\ref{fig:sdss_panels}, with the exception of the rightmost column. Since there is no direct counterpart to the Coma cluster in the SAM, we instead apply an analogous cut by removing centrals near to haloes with $M_{\mathrm{vir}}>10^{14.4}\Msun$ (i.e. $R_{\mathrm{vir}}>2\,\mathrm{Mpc}$), by applying the same 4 Mpc cylinder cut as described in Section~\ref{sec:re_analysis}. Note that the error-bars are very much smaller in Fig.~\ref{fig:mock_panels} than in Fig.~\ref{fig:sdss_panels} because of the 240-fold increase in the number of objects in the mock.

Before examining these results, some key differences between the SAM and the SDSS data should be noted. First, the Main Sequence in the SAM is more sharply peaked than in the data, and this is the cause of the systematic vertical (sSFR) offset between the results for the medians in the top two rows of Fig.~\ref{fig:sdss_panels} and \ref{fig:mock_panels}. The offset is much less pronounced for the mean sSFR in the lower two rows. Second, unlike in the real data, most of the galaxies in the passive population in the SAM have exactly zero SFR. In our treatment of the SAM mock catalogue, galaxies with sSFRs below a threshold of $10^{-12} \,\mathrm{yr^{-1}}$ are assigned sSFRs that are randomly drawn from a Gaussian distribution centered on $\sim 10^{-11.6} \,\mathrm{yr^{-1}}$ with a dispersion of 0.5 dex (see Section~\ref{subsec:mock_cat}), so that the sSFR-mass distribution in the SAM approximately matches that of the observations. Because of this scrambling of sSFRs for low-sSFR centrals, we treat centrals in the lowest two quartiles as a single set with a single combined neighbour sample. As a consequence of this, we plot on Fig.~\ref{fig:mock_panels} only a single set of points representing the two lowest quartiles combined, which may of course dilute the signal that would have been obtained if the lowest quartile could have been studied in isolation. It should also be noted that the lowest two sSFR-quartiles in Fig.~\ref{fig:fig_purity} and \ref{fig:2d_purity} are artificially similar for the same reason. 

These small systematic differences aside, the results from the SAM in Fig.~\ref{fig:mock_panels} are strikingly similar to the observational results shown in Fig.~\ref{fig:sdss_panels}. With only the prescriptions of \lq known\rq\ physics in the model, the application of the K13 methodology nevertheless yields the appearance of large-scale sSFR correlations that are similar to those observed in the SDSS data. The similarity between the results from the real observational data and from the simulated data, in terms of both the amplitude and the range of the sSFR correlation, suggests that there is no need (at least from this analysis) to add new physics to our view of galaxy evolution. 

We note that we have also analysed the original \citet{Guo2011} SAM mock catalogue used in K13 in the same way. We find qualitatively similar results as in Fig.~\ref{fig:mock_panels}, and therefore cannot account for the apparent absence of a similar signal in K13\rq s treatment of this same mock catalogue.

However, we note that the change in the conformity signal is not identical between the two sets of data for all of the modifications. In particular, the percentage of centrals removed from the mock sample at each cut is systematically higher than in the SDSS data, and the effect of removing centrals in rich environments (i.e. operations (i) and (ii) as described in Section~\ref{sec:re_analysis}) produces a less dramatic reduction of the signal in the mock data compared with the SDSS data.

We also note that, even when accounting for all of the methodological issues, there exists a weak conformity signal in the mock that is not seen in the data. Since inter-halo interactions are not present in the SAM, this signal must be due to the spatial correlation of halo properties, such as those mentioned in the beginning of this Section.

These differences are not entirely surprising, considering that the SDSS volume is approximately 200 times smaller than the simulation volume of the H15 SAM. In order to understand how much of these differences are simply due to the limited volume of the SDSS sample, we select, from the mock, 125 independent sub-volumes which have similar spatial dimensions to our SDSS sample, and examine the variation in the sSFR distribution of the neighbours.

Fig.~\ref{fig:realn_noise} shows the result of this analysis. The top row shows ${\sigma_{\mathrm{Real.}}}^{2}$, the variance of the median neighbour sSFR across the 125 independent realizations, and reflects the total variance in an SDSS-like volume. The middle row shows $\langle{\sigma_{\mathrm{Boot.}}}^{2}\rangle$, the median value of the bootstrap variances, where the median averages over all of the realizations. In this analysis, $\langle{\sigma_{\mathrm{Boot.}}}^{2}\rangle$ therefore reflects the estimate of Poisson uncertainties due to the sample size. Finally, the bottom row shows the square root of the ratio of these two quantities, and effectively indicates how well the bootstrap error-bars (i.e. in Fig. \ref{fig:confor_kauff}, \ref{fig:sdss_panels}, and \ref{fig:rem_cyl}) reflect the total variance.

The Poisson variance of neighbour sSFRs, i.e. the variance estimated from bootstrap resampling, is dominant at small radii, which corresponds to the dominance of uncertainties due to low number of neighbours per radial bin. At larger radii, the effects of volume-limited realization on the variance, i.e. \lq cosmic variance\rq, increase in relative importance, especially for the low-sSFR centrals and neighbours. This is due to the fact that, as we have shown, the neighbour sSFR distribution is influenced strongly by the presence of a small number of large nearby structures. The bootstrap resampling of the neighbours does not capture the small-number statistics of these rich clusters. As a result, the bootstrap error-bars in Fig.~\ref{fig:confor_kauff}, \ref{fig:sdss_panels}, and \ref{fig:rem_cyl} underestimate the true uncertainty in the sSFR distribution of neighbours by at least a factor of 2. This should be borne in mind when comparing the observational data with the mock catalogue.

Therefore, we suspect that the offset between the results from the SDSS data and from the mock data is not statistically significant. The similarity of the results under the various methodological modifications confirms our assertion that the bulk of the conformity signal is indeed driven by a known (and accounted-for) correlation which has been amplified by the density-weighting effect.

Since the bootstrap uncertainties for the current SDSS sample underestimate the true uncertainty, the apparent absence of a conformity signal after the methodological modifications (i.e. in the $4^{\rm{th}}$ row, $2^{\rm{nd}}$ column of Fig.~\ref{fig:sdss_panels}) does not rule out the existence of a weak signal at the level seen in the mock.

\section{Discussion}
\label{sec:discussion}
The main conclusion from our analysis is that a number of methodological issues can substantially amplify the strength of sSFR correlations. Some of these are obvious, such as the use of median statistics for a bimodal distribution, and some are more subtle, including the effective density-weighting of the pair-counting scheme in K13. There is also the issue of central-satellite misclassification, although we stress that removing all satellites (as identified in the group catalogue) does not completely eliminate the 4 Mpc conformity signal in our analysis. 

It is then clear from our analysis that the 4 Mpc scale conformity signal is actually being driven (via these amplification effects) by a very small number of centrals that live on the outskirts of the largest clusters in the Universe. The fact that their sSFRs are correlated with those of the large number of galaxies in the clusters appears to be a real effect, albeit greatly boosted in K13 by the methodological aspects discussed in the current paper. 

In the final stages of manuscript preparation of this paper, \citet{Tinker2017} posted a pre-print in which they reproduced the K13 result (i.e. Fig.~\ref{fig:confor_kauff}) by using the same methodology as K13. They also identify probable satellite contaminants (which they refer to as \lq non-pure\rq\ centrals) in the sample of centrals by using their group catalogue, and found that the large-scale conformity signal is effectively eliminated when probable contaminants are excluded. 

Through private correspondence, we found that among the centrals which they classify as \lq non-pure\rq, some fall under our category of \lq rich\rq\ ($\rm{N}_{neigh}\,>$ 70) centrals. This intersecting subsample makes up $\sim$\,20\% of their \lq non-pure\rq\ sample, and $\sim$\,25\% of our \lq rich\rq\ sample. The same reduction in the conformity signal (in their fig. 5) can be achieved by only removing this intersecting subset of \lq rich non-pure\rq\ centrals, while the removal of \lq non-pure\rq\ centrals with fewer than 70 neighbours has essentially no impact on the conformity signal. This is consistent with our identification of the origin of the K13 conformity signal, namely that the strong effect is primarily driven by centrals in high-density regions, and is further amplified by satellite contaminants.

Similar methodological issues have also been addressed in analyses of other data sets, with varying results. \citet{Berti2017} investigated conformity in a sample of PRIMUS galaxies at intermediate redshift. The authors did this by splitting centrals into passive and star-forming populations, and quantified the star-formation of neighbour galaxies using their star-forming fraction. The authors explicitly tested the impact of the different weighting schemes of centrals, and found that the conformity signal, as quantified by the star-forming fraction, was insensitive to the choice of weighting scheme. However, we note that under the \lq density-weighted\rq\ scheme, the statistical significance of their measured large-scale conformity signal is far smaller than that in K13. It is therefore unclear whether this insensitivity was due to the fact that the star-forming fraction is a more robust marker of conformity, or simply that the conformity signal is intrinsically weaker in that sample.

The effects of satellite contamination in the selection of centrals were discussed in \citet{Bray2016}, where the authors investigated conformity in the Illustris simulation. They found that while satellite contamination in the \lq central\rq\ sample does indeed amplify the observed conformity signal, a weak large-scale conformity signal out to $\sim$ 3\,Mpc can be detected in the simulation even after the satellite contaminants are removed.

In both cases, a large-scale conformity signal is detected out to $\sim$ 3\,Mpc even when accounting for the highlighted methodological issues. That is to say, while the methodological issues greatly amplify the conformity signal in high-density regions, there may be a weaker, true, large-scale conformity signal in the Universe. Since this work, following K13, investigates conformity using the full sSFR distribution, and not just the star-forming fraction, it is difficult to compare the strengths of the underlying conformity signals between these works. However, qualitatively, this correlation appears to be present in the SAM mock catalogues, and could have a number of origins, including effects like assembly history bias or other environment-based effects that do not involve direct super-halo interactions (since these are not in the SAM).

The clear identification in this paper that the effect is being driven by low-sSFR \lq centrals\rq\ (some of which are probably real centrals, although some are likely misidentified satellites) in the close vicinity of very massive clusters (which is clear from Fig.~\ref{fig:ssfr_hist}, \ref{fig:ra_dec_map}, \ref{fig:sdss_panels}, and \ref{fig:rem_cyl}) emphasizes the difficulty of correctly interpreting the \lq scale\rq\ of the conformity signal. Rather than a \lq long-range\rq\ effect, operating at about 10 virial radii from the small haloes hosting the set of centrals, it should better be thought of as a \lq short-range\rq\ effect, operating at about one virial radius from the very large haloes that are hosting the neighbours of those centrals. 

In a more general sense, this also highlights the importance of matching the centrals and neighbours in conformity studies, as discussed at length in \citet{Knobel2015}. Although the centrals with high and low sSFR are reasonably well-matched in stellar mass in K13, and likely therefore also in their own halo mass, it is clear (from Fig.~\ref{fig:ssfr_hist}) that they do not inhabit the same range of Mpc-scale environments. In particular, it is clear from Fig.~\ref{fig:ssfr_hist} that the set of low-sSFR centrals inhabit a much broader range of environments than the set of high-sSFR centrals, and that the (relatively few) centrals with the most neighbours ($N_{\mathrm{neigh}} > 70$ within 4 Mpc) are predominantly of low sSFR. 

This is the origin of the conformity signal: it seems much more plausible that the signal comes from the very special environment where these few richest centrals lie, rather than from the centrals themselves.

\section{Summary and conclusions}
\label{sec:summary}
This paper has re-examined the observational evidence in the SDSS for galactic conformity effects at large scales, as presented in \citet[][K13]{Kauffmann2013}. For simplicity, we focused on the analysis of the set of centrals of intermediate stellar mass $10^{10} \Msun< M_* < 10^{10.5}\Msun$, where the conformity signal in K13 is strongest. Likewise, we considered only the simple (total) sSFR as the indicator of star-formation activity. 

We first identify three features of the K13 analysis methodology that we have shown to artificially introduce or amplify a conformity signal:

\begin{enumerate}
\item{The K13 analysis is implicitly weighted towards those central galaxies which have large numbers of neighbours. Since these centrals have both generally low sSFR and have low-sSFR neighbours, this produces a positive conformity signal. The preferential weighting of these centrals boosts (in proportion to their number of neighbours) their contribution to the overall conformity signal in the sample.}

\item{Some centrals selected by the K13 isolation criterion are likely to be misclassified satellite galaxies. This can produce a spurious conformity signal if the rate of misclassification is correlated with the overall passive fraction of the satellites, which it appears to be. Since the probability of misclassification also appears to increase with the number of neighbours, the weighting of the sample in favour of centrals with many neighbours further exacerbates this problem.}

 \item{In addition, the use of the median to describe the sSFR distribution of the neighbour galaxies further amplifies the size of the conformity signal. Since the neighbour galaxies have a bimodal distribution of sSFR, with roughly equal strengths of the two components, a small shift in the relative numbers of high- and low-sSFR neighbours results in a large change in the median, about twice the change in the mean (of the logarithm).}
\end{enumerate}

We then re-analyse the SDSS data with various combinations of small but significant modifications to the analysis methodology based on these three issues. The combination of these modifications dramatically reduces the large-scale conformity signal to the level that it can no longer be detected with the available data.

Removing the weighting in favour of centrals with many neighbours is already sufficient to vastly reduce the conformity signal, to the extent that the amplitude of the remaining signal is comparable to the size of the estimated uncertainties.

Even without removing the implicit density-weighting, the signal beyond 2 Mpc essentially disappears if the 18 centrals within 4 Mpc of the Coma cluster are removed. More than half of these \lq rich centrals\rq\ are likely to be misclassified satellites, but some may well be real centrals. These centrals are preferentially of low sSFR, and the large number of neighbours are also preferentially of low sSFR, thereby producing a conformity signal. This signal is only present for a very small number of rich centrals in the vicinity of a few large clusters, but it came to produce the large overall effect in the K13 results via the density-weighting that is implicit in their method.

This result emphasizes the difficulty of correctly interpreting the \lq scale\rq\ of the conformity signal. While a 4 Mpc-scale correlation may appear to be an extremely \lq long-range\rq\ effect when compared with the virial sizes of the relatively low-mass centrals under consideration, we have illustrated that it is an effect that arises within approximately one virial radius of the largest haloes. Indeed, we show that progressively removing centrals in the vicinity of large clusters systematically reduces the spatial extent of the large-scale conformity signal. The large-scale conformity effect seen in K13 should therefore better be thought of as a \lq short-range\rq\ effect, associated with the environmental quenching effects of neighbours around the larger haloes, rather than a very \lq long-range\rq\ effect driven by the smaller haloes.

Finally, we also analyse the mock catalogue from the \citet{Henriques2015} semi-analytic model in exactly the same way as the SDSS data. Both the effects of the methodological issues, and also the overall levels of conformity seen at each step, are very similar in the real and mock data, suggesting little need for the inclusion of any new physical processes in the models in order to address large-scale conformity.

Because the signal is dominated by the centrals that are located in the neighbourhood of a handful of the richest clusters, the actual uncertainties are substantially larger than those estimated by the bootstrap resampling. This should be borne in mind when comparing results from the real data with those from the mock catalogue.

\section*{Acknowledgements}
We thank Joanna Woo for kindly providing the original compilation of the SDSS catalogues, and also Christian Knobel and Aseem Paranjape for previous discussions on galactic conformity. This work has been supported by the Swiss National Science Foundation. 
BMBH (ORCID 0000-0002-1392-489X) acknowledges support from an ETH Zwicky Prize Fellowship.

\bibliographystyle{mn2e} \bibliography{paper}

\begin{thebibliography}{}

\bibitem[\protect\citeauthoryear{{Abadi}, {Moore} \& {Bower}}{{Abadi}
  et~al.}{1999}]{Abadi1999}
{Abadi} M.~G.,  {Moore} B.,    {Bower} R.~G.,  1999, MNRAS, 308, 947

\bibitem[\protect\citeauthoryear{{Abazajian} et~al.,}{{Abazajian}
  et~al.}{2009}]{Abazajian2009}
{Abazajian} K.~N.,  et~al., 2009, ApJS, 182, 543

\bibitem[\protect\citeauthoryear{{Baldry}, {Balogh}, {Bower}, {Glazebrook},
  {Nichol}, {Bamford} \& {Budavari}}{{Baldry} et~al.}{2006}]{Baldry2006}
{Baldry} I.~K.,  {Balogh} M.~L.,  {Bower} R.~G.,  {Glazebrook} K.,  {Nichol}
  R.~C.,  {Bamford} S.~P.,    {Budavari} T.,  2006, MNRAS, 373, 469

\bibitem[\protect\citeauthoryear{{Baldry}, {Glazebrook}, {Brinkmann},
  {Ivezi{\'c}}, {Lupton}, {Nichol} \& {Szalay}}{{Baldry}
  et~al.}{2004}]{Baldry2004}
{Baldry} I.~K.,  {Glazebrook} K.,  {Brinkmann} J.,  {Ivezi{\'c}} {\v Z}.,
  {Lupton} R.~H.,  {Nichol} R.~C.,    {Szalay} A.~S.,  2004, ApJ, 600, 681

\bibitem[\protect\citeauthoryear{{Berti}, {Coil}, {Behroozi}, {Eisenstein},
  {Bray}, {Cool} \& {Moustakas}}{{Berti} et~al.}{2017}]{Berti2017}
{Berti} A.~M.,  {Coil} A.~L.,  {Behroozi} P.~S.,  {Eisenstein} D.~J.,  {Bray}
  A.~D.,  {Cool} R.~J.,    {Moustakas} J.,  2017, ApJ, 834, 87

\bibitem[\protect\citeauthoryear{{Blanton} et~al.,}{{Blanton}
  et~al.}{2005}]{Blanton2005}
{Blanton} M.~R.,  et~al., 2005, AJ, 129, 2562

\bibitem[\protect\citeauthoryear{{Bluck}, {Mendel}, {Ellison}, {Moreno},
  {Simard}, {Patton} \& {Starkenburg}}{{Bluck} et~al.}{2014}]{Bluck2014}
{Bluck} A.~F.~L.,  {Mendel} J.~T.,  {Ellison} S.~L.,  {Moreno} J.,  {Simard}
  L.,  {Patton} D.~R.,    {Starkenburg} E.,  2014, MNRAS, 441, 599

\bibitem[\protect\citeauthoryear{{Bower}, {Benson}, {Malbon}, {Helly}, {Frenk},
  {Baugh}, {Cole} \& {Lacey}}{{Bower} et~al.}{2006}]{Bower2006}
{Bower} R.~G.,  {Benson} A.~J.,  {Malbon} R.,  {Helly} J.~C.,  {Frenk} C.~S.,
  {Baugh} C.~M.,  {Cole} S.,    {Lacey} C.~G.,  2006, MNRAS, 370, 645

\bibitem[\protect\citeauthoryear{{Bray} et~al.,}{{Bray}
  et~al.}{2016}]{Bray2016}
{Bray} A.~D.,  et~al., 2016, MNRAS, 455, 185

\bibitem[\protect\citeauthoryear{{Brinchmann}, {Charlot}, {White}, {Tremonti},
  {Kauffmann}, {Heckman} \& {Brinkmann}}{{Brinchmann}
  et~al.}{2004}]{Brinchmann2004}
{Brinchmann} J.,  {Charlot} S.,  {White} S.~D.~M.,  {Tremonti} C.,  {Kauffmann}
  G.,  {Heckman} T.,    {Brinkmann} J.,  2004, MNRAS, 351, 1151

\bibitem[\protect\citeauthoryear{{Croton} et~al.,}{{Croton}
  et~al.}{2006}]{Croton2006}
{Croton} D.~J.,  et~al., 2006, MNRAS, 365, 11

\bibitem[\protect\citeauthoryear{{Daddi} et~al.,}{{Daddi}
  et~al.}{2007}]{Daddi2007}
{Daddi} E.,  et~al., 2007, ApJ, 670, 156

\bibitem[\protect\citeauthoryear{{Elbaz} et~al.,}{{Elbaz}
  et~al.}{2007}]{Elbaz2007}
{Elbaz} D.,  et~al., 2007, Astronomy and Astrophysics Supplement Series, 468,
  33

\bibitem[\protect\citeauthoryear{{Gao}, {Springel} \& {White}}{{Gao}
  et~al.}{2005}]{Gao2005}
{Gao} L.,  {Springel} V.,    {White} S.~D.~M.,  2005, MNRAS, 363, L66

\bibitem[\protect\citeauthoryear{{Gunn} \& {Tinsley}}{{Gunn} \&
  {Tinsley}}{1976}]{Gunn1976}
{Gunn} J.~E.,  {Tinsley} B.~M.,  1976, ApJ, 210, 1

\bibitem[\protect\citeauthoryear{{Guo} et~al.,}{{Guo}  et~al.}{2011}]{Guo2011}
{Guo} Q.,  et~al., 2011, MNRAS, 413, 101

\bibitem[\protect\citeauthoryear{{Hearin}, {Behroozi} \& {van den
  Bosch}}{{Hearin} et~al.}{2016}]{Hearin2016}
{Hearin} A.~P.,  {Behroozi} P.~S.,    {van den Bosch} F.~C.,  2016, MNRAS, 461,
  2135

\bibitem[\protect\citeauthoryear{{Hearin}, {Watson} \& {van den
  Bosch}}{{Hearin} et~al.}{2015}]{Hearin2015}
{Hearin} A.~P.,  {Watson} D.~F.,    {van den Bosch} F.~C.,  2015, MNRAS, 452,
  1958

\bibitem[\protect\citeauthoryear{{Henriques}, {White}, {Thomas}, {Angulo},
  {Guo}, {Lemson}, {Springel} \& {Overzier}}{{Henriques}
  et~al.}{2015}]{Henriques2015}
{Henriques} B.~M.~B.,  {White} S.~D.~M.,  {Thomas} P.~A.,  {Angulo} R.,  {Guo}
  Q.,  {Lemson} G.,  {Springel} V.,    {Overzier} R.,  2015, MNRAS, 451, 2663

\bibitem[\protect\citeauthoryear{{Henriques}, {White}, {Thomas}, {Angulo},
  {Guo}, {Lemson} \& {Wang}}{{Henriques} et~al.}{2016}]{Henriques2016}
{Henriques} B.~M.~B.,  {White} S.~D.~M.,  {Thomas} P.~A.,  {Angulo} R.~E.,
  {Guo} Q.,  {Lemson} G.,    {Wang} W.,  2016, ArXiv e-prints

\bibitem[\protect\citeauthoryear{{Kauffmann}}{{Kauffmann}}{2015}]{Kauffmann2015}
{Kauffmann} G.,  2015, MNRAS, 454, 1840

\bibitem[\protect\citeauthoryear{{Kauffmann} et~al.,}{{Kauffmann}
  et~al.}{2003}]{Kauffmann2003}
{Kauffmann} G.,  et~al., 2003, MNRAS, 341, 33

\bibitem[\protect\citeauthoryear{{Kauffmann}, {Li}, {Zhang} \&
  {Weinmann}}{{Kauffmann} et~al.}{2013}]{Kauffmann2013}
{Kauffmann} G.,  {Li} C.,  {Zhang} W.,    {Weinmann} S.,  2013, MNRAS, 430,
  1447

\bibitem[\protect\citeauthoryear{{Knobel}, {Lilly}, {Woo} \& {Kova{\v
  c}}}{{Knobel} et~al.}{2015}]{Knobel2015}
{Knobel} C.,  {Lilly} S.~J.,  {Woo} J.,    {Kova{\v c}} K.,  2015, ApJ, 800, 24

\bibitem[\protect\citeauthoryear{{Kubo}, {Stebbins}, {Annis}, {Dell'Antonio},
  {Lin}, {Khiabanian} \& {Frieman}}{{Kubo} et~al.}{2007}]{Kubo2007}
{Kubo} J.~M.,  {Stebbins} A.,  {Annis} J.,  {Dell'Antonio} I.~P.,  {Lin} H.,
  {Khiabanian} H.,    {Frieman} J.~A.,  2007, ApJ, 671, 1466

\bibitem[\protect\citeauthoryear{{Larson}, {Tinsley} \& {Caldwell}}{{Larson}
  et~al.}{1980}]{Larson1980}
{Larson} R.~B.,  {Tinsley} B.~M.,    {Caldwell} C.~N.,  1980, ApJ, 237, 692

\bibitem[\protect\citeauthoryear{{Moore}, {Katz}, {Lake}, {Dressler} \&
  {Oemler}}{{Moore} et~al.}{1996}]{Moore1996}
{Moore} B.,  {Katz} N.,  {Lake} G.,  {Dressler} A.,    {Oemler} A.,  1996,
  Nature, 379, 613

\bibitem[\protect\citeauthoryear{{Noeske} et~al.,}{{Noeske}
  et~al.}{2007}]{Noeske2007}
{Noeske} K.~G.,  et~al., 2007, ApJ, 660, L43

\bibitem[\protect\citeauthoryear{{Paranjape}, {Kova{\v c}}, {Hartley} \&
  {Pahwa}}{{Paranjape} et~al.}{2015}]{Paranjape2015}
{Paranjape} A.,  {Kova{\v c}} K.,  {Hartley} W.~G.,    {Pahwa} I.,  2015,
  MNRAS, 454, 3030

\bibitem[\protect\citeauthoryear{{Peng} et~al.,}{{Peng}
  et~al.}{2010}]{Peng2010}
{Peng} Y.-j.,  et~al., 2010, ApJ, 721, 193

\bibitem[\protect\citeauthoryear{{Peng}, {Lilly}, {Renzini} \&
  {Carollo}}{{Peng} et~al.}{2012}]{Peng2012}
{Peng} Y.-j.,  {Lilly} S.~J.,  {Renzini} A.,    {Carollo} M.,  2012, ApJ, 757,
  4

\bibitem[\protect\citeauthoryear{{Planck Collaboration}}{{Planck
  Collaboration}}{2014}]{Planck2014}
{Planck Collaboration} 2014, Astronomy and Astrophysics Supplement Series, 571,
  A16

\bibitem[\protect\citeauthoryear{{Salim} et~al.,}{{Salim}
  et~al.}{2007}]{Salim2007}
{Salim} S.,  et~al., 2007, ApJS, 173, 267

\bibitem[\protect\citeauthoryear{{Tinker}, {Hahn}, {Mao}, {Wetzel} \&
  {Conroy}}{{Tinker} et~al.}{2017}]{Tinker2017}
{Tinker} J.~L.,  {Hahn} C.,  {Mao} Y.-Y.,  {Wetzel} A.~R.,    {Conroy} C.,
  2017, ArXiv e-prints

\bibitem[\protect\citeauthoryear{{van den Bosch}, {Aquino}, {Yang}, {Mo},
  {Pasquali}, {McIntosh}, {Weinmann} \& {Kang}}{{van den Bosch}
  et~al.}{2008}]{vdBosch2008}
{van den Bosch} F.~C.,  {Aquino} D.,  {Yang} X.,  {Mo} H.~J.,  {Pasquali} A.,
  {McIntosh} D.~H.,  {Weinmann} S.~M.,    {Kang} X.,  2008, MNRAS, 387, 79

\bibitem[\protect\citeauthoryear{{Weinmann}, {van den Bosch}, {Yang} \&
  {Mo}}{{Weinmann} et~al.}{2006}]{Weinmann2006}
{Weinmann} S.~M.,  {van den Bosch} F.~C.,  {Yang} X.,    {Mo} H.~J.,  2006,
  MNRAS, 366, 2

\bibitem[\protect\citeauthoryear{{Woo}, {Dekel}, {Faber} \& {Koo}}{{Woo}
  et~al.}{2015}]{Woo2015}
{Woo} J.,  {Dekel} A.,  {Faber} S.~M.,    {Koo} D.~C.,  2015, MNRAS, 448, 237

\bibitem[\protect\citeauthoryear{{Yang}, {Mo} \& {van den Bosch}}{{Yang}
  et~al.}{2006}]{Yang2006}
{Yang} X.,  {Mo} H.~J.,    {van den Bosch} F.~C.,  2006, ApJ, 638, L55

\bibitem[\protect\citeauthoryear{{Yang}, {Mo}, {van den Bosch}, {Pasquali},
  {Li} \& {Barden}}{{Yang} et~al.}{2007}]{Yang2007}
{Yang} X.,  {Mo} H.~J.,  {van den Bosch} F.~C.,  {Pasquali} A.,  {Li} C.,
  {Barden} M.,  2007, ApJ, 671, 153

\end{thebibliography}

\label{lastpage}

\end{document}